\documentclass[conference]{IEEEtran}
\usepackage{cite}
\usepackage{amsmath,amssymb,amsfonts}
\usepackage{algorithmic}
\usepackage{graphicx}
\usepackage{textcomp}
\usepackage{xcolor}
\usepackage[hyphens]{url}
\usepackage{amsmath}
\usepackage{hhline}
\usepackage{silence}
\usepackage{graphicx}
\usepackage{multirow} 
\usepackage{graphicx}
\usepackage{multirow}
\usepackage{amsmath}
\usepackage[english]{babel}
\usepackage{booktabs}
\usepackage{array}
\usepackage{paralist}
\usepackage{threeparttable}
\usepackage{lipsum}
\usepackage{flushend}
\usepackage{cuted}
\usepackage{amssymb}
\usepackage{color}
\usepackage{psfrag}
\usepackage{epsfig}
\usepackage{multicol}
\usepackage{color}
\usepackage{fancybox}
\usepackage{cite}
\usepackage{theorem}
\usepackage{url}
\usepackage[normalem]{ulem}
\usepackage{siunitx}
\usepackage{caption}
\usepackage{subfig}
\usepackage[font=footnotesize,labelfont=bf]{caption}
\usepackage{silence}
\usepackage{enumitem}
\definecolor{gray1}{gray}{0.7}
\definecolor{gray2}{gray}{0.98}
\definecolor{light-gray}{gray}{0.95}

\newcommand{\ignore}[1]{}

\newcommand{\blueHL}[1]{{\textcolor{blue}{#1}}}

\usepackage{tikz}
\usetikzlibrary{shapes,arrows}
\usepackage{flexisym}

\def\BibTeX{{\rm B\kern-.05em{\sc i\kern-.025em b}\kern-.08em
    T\kern-.1667em\lower.7ex\hbox{E}\kern-.125emX}}

\title{GNNIE: GNN Inference Engine \\
with Load-balancing and Graph-Specific Caching} 
\author{
\vspace*{-0.3in}\\
Sudipta Mondal, Susmita Dey Manasi, Kishor Kunal, S. Ramprasath, and Sachin S. Sapatnekar\\
ECE Department, University of Minnesota\\
\vspace*{-0.3in}
}

\begin{document}
\maketitle
\thispagestyle{plain}
\pagestyle{plain}


\begin{abstract}
Graph neural networks (GNN) analysis engines are vital for real-world
problems that use large graph models.  Challenges for a GNN
hardware platform include the ability to (a) host a variety of GNNs, (b) handle
high sparsity in input vertex feature vectors and the graph adjacency matrix
and the accompanying random memory access patterns, and (c) maintain
load-balanced computation in the face of uneven workloads, induced by high
sparsity and power-law vertex degree distributions.  This paper proposes GNNIE,
an accelerator designed to run a broad range of GNNs.  It tackles workload
imbalance by (i)~splitting vertex feature operands into blocks, (ii)~reordering
and redistributing computations, (iii)~using a novel flexible MAC architecture.
It adopts a graph-specific, degree-aware caching policy that is well
suited to real-world graph characteristics. The policy enhances on-chip data
reuse and avoids random memory access to DRAM.

GNNIE achieves average speedups of 21233$\times$ over a CPU and 699$\times$
over a GPU over multiple datasets on graph attention networks (GATs), graph
convolutional networks (GCNs), GraphSAGE, GINConv, and DiffPool. Compared to
prior approaches, GNNIE achieves an average speedup of 35$\times$ over HyGCN
(which cannot implement GATs) for GCN, GraphSAGE, and GINConv, and, using
3.4$\times$ fewer processing units, an average speedup of 2.1$\times$ over
AWB-GCN (which runs only GCNs).
\end{abstract}

\vspace{-1mm}
\section{Introduction}
\label{sec:Intro}

\noindent 
Deep learning accelerators have largely focused on data with Euclidean
embeddings, e.g., audio/video/images/speech.  Many
real-world problems (e.g., network analysis, embedded sensing, e-commerce,
drug interactions) use graphs to model relationships. Inferencing on large,
unstructured, and sparse graphs with non-Euclidean embeddings
requires specialized graph neural networks (GNNs).
Today's GNNs~\cite{kipf2016semi, hamilton2017inductive, velivckovic2017graph_2,
xu2019ginconv} are based on nearest-neighbor operations, with lower computation
cost than early approaches \cite{Bruna2013, Defferrard2016, kipf2016semi}.
Though GNNs can be run on software
platforms~\cite{yang2019aligraph,wang2020deep,tensorflow}, high performance
requires hardware acceleration platforms.

Multilayer GNNs perform two computation steps per layer:\\
\textbf{\em (a) Weighting} performs a linear
transform of vertex feature vectors through multiplication by a weight matrix.
\\ \textbf{\em (b) Aggregation} consolidates information from the neighbors of
a vertex to compute the feature vectors for the next layer.  

\begin{figure}[!htb]
\centering
\includegraphics[width=2.2in]{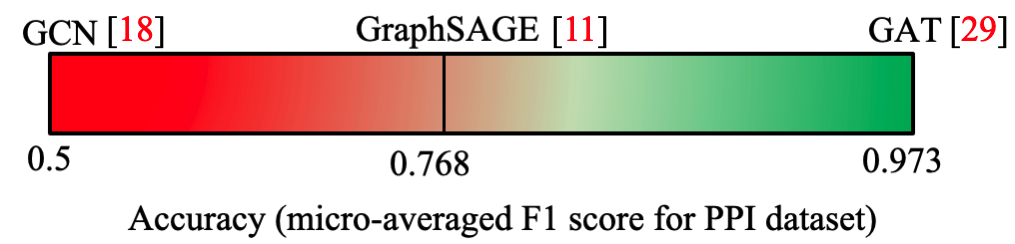}
\vspace{-2mm}
\caption{{GNN accuracy comparison (data from~\cite{velivckovic2017graph_2}, PPI dataset).}
}
\label{fig:GNN accuracy comparison}
\vspace{-6mm}
\end{figure}

\noindent
The challenges in building GNN accelerators are as follows:

\noindent
{\em (1) Versatility}
An accelerator should be able to handle a diverse set of GNNs to cover a wide
range of GNN architectures to provide appropriate computation/accuracy tradeoff
points for various applications.  The achievable accuracy depends on the GNN:
graph attention networks (GATs) achieve higher accuracy than other GNNs, but 
with more computation (Fig.~\ref{fig:GNN accuracy comparison}).

\noindent
{\em (2) Adjacency matrix sparsity}
The graph adjacency matrix encodes vertex neighborhood information
required for Aggregation.  The adjacency matrix is {\em highly
sparse} ($>99.8\%$ for all datasets in this paper; in contrast, DNN data shows
10\%--50\% sparsity).  Unlike image/video data, adjacency matrix
sparsity patterns typically exhibit {\em power-law behavior}, with
vertex degrees ranging from very low (for most vertices) to extremely high
(for very few vertices): in the Reddit dataset, $11\%$ of the vertices
cover $88\%$ of all edges.  

\begin{figure}[hbt]
\vspace*{-1mm}
\centering
\vspace{-2mm}
\includegraphics[width=2.1in]{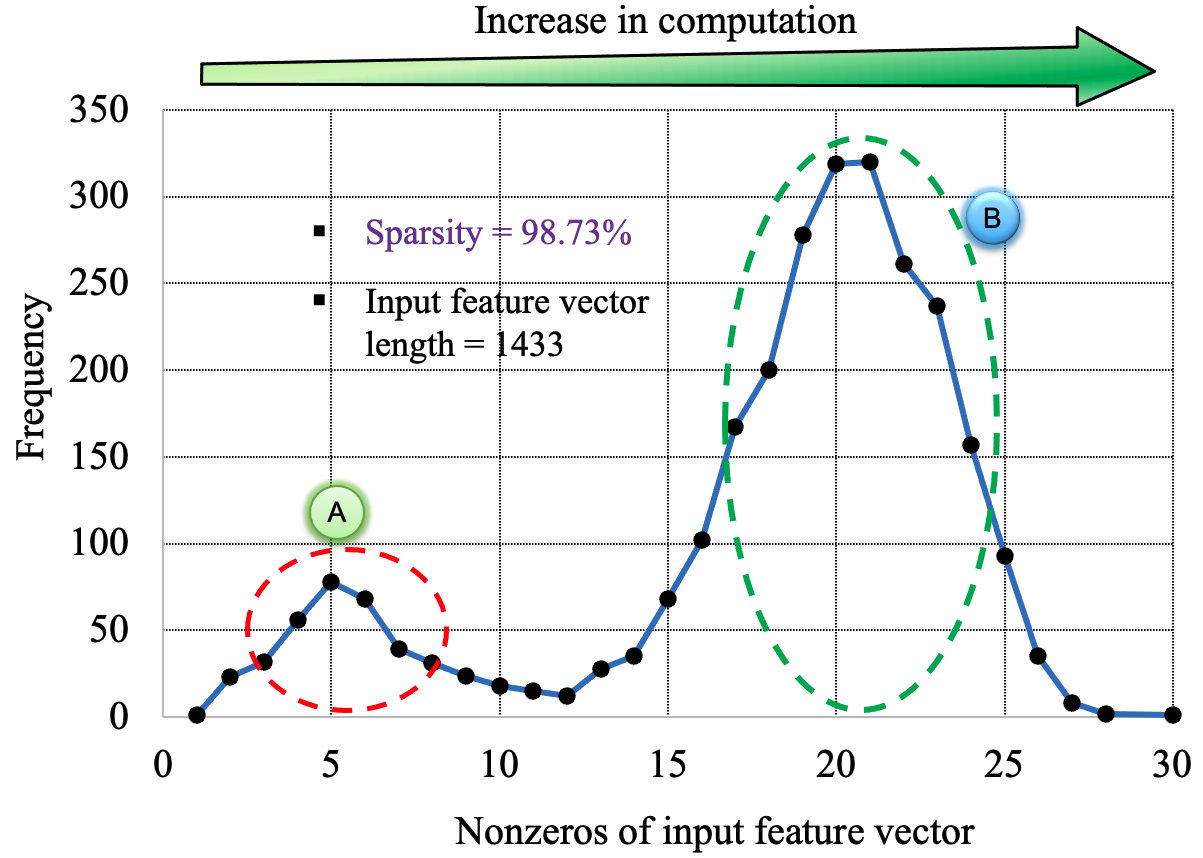}
\vspace{-2mm}
\caption{Nonzero histogram for input vertex feature vectors (Cora).}
\label{fig:input feature vector sparsity}
\vspace{-3mm}
\end{figure}

\noindent
{\em (3) Input feature vector sparsity}
The vertex {\em input feature vectors} are highly sparse, e.g., the 2708 input
vertex feature vectors of the Cora dataset have $98.73\%$ average sparsity. 
In Fig.~\ref{fig:input feature vector sparsity},  Region~A is sparser than B
and requires less computation, leading to load balancing issues during
Weighting.

\noindent
{\em (4) Memory footprint and random-access patterns}
Real-world graphs have a large number of vertices and a massive memory
footprint (Reddit: 2.3Gb in sparse format).  High sparsity and power-law 
distributions can lead to random memory access patterns and poor data access
locality in Aggregation.  

Therefore, GNN-specific accelerators must address:\\
(a)~{\em load balancing during Weighting}, due to the sparsity variations in
Fig.~\ref{fig:input feature vector sparsity}, and {\em during Aggregation}, due to
the imbalance of computations for high- and low-degree vertices.
(b)~{\em lightweight graph-specific caching} of the adjacency matrix for high
data access locality and maximal reuse of cached data.

\noindent
{\bf Relation to other acceleration engines:}
The Weighting step performs matrix-vector multiplication which resembles convolutional neural network (CNN)
computations, but CNN accelerators~\cite{EIE2016, chen2016eyeriss,
aimar2018nullhop, Jouppi2017, SCNN2017, BitFusion2018, Jouppi2020} are
inefficient at handling graph data.  Aggregation operates on
graph neighborhoods and resembles graph analytics, but graph processing
accelerators~\cite{ham2016graphicionado,Dai16,jun2018grafboost} are designed to
perform lightweight computations, significantly lower than the needs of a GNN.
Extensions of CNN/graph processing engines are inadequate.

An early GNN accelerator, HyGCN~\cite{yan2020hygcn}, bridges the divide by
using two pipelined engines: an Aggregation engine that operates on
graph data and consolidates vertex feature vectors from the neighborhood of
each vertex, followed by a Combination engine, which uses a multilevel
perceptron to weight the aggregated features with the weight matrix.  
The disparity between engines raises challenges in providing a steady stream of
data to keep the Aggregation/Combination engine pipeline busy. The Aggregation
engine does not account for power-law behavior while caching partial results,
and high-degree vertices may create stalls due to the limited size of on-chip
buffers.  In the Combination engine, the aggregated feature vectors
are both sparse and show high sparsity variations (Fig.~\ref{fig:input feature
vector sparsity}).  Consequently, stalls are required, leading to inefficiency.

AWB-GCN~\cite{Geng2020AWBGCNAG} views the GNN computation as two consecutive
sparse-dense matrix multiplications (SpMMs).  During Weighting, the method is
targeted to moderate sparsity of 75\% -- but input layer vertex feature vectors
are ultra-sparse (Fig.~\ref{fig:input feature vector sparsity}).  During
Aggregation, the graph-agnostic SpMM view necessitates numerous expensive
off-chip accesses to the adjacency matrix.  AWB-GCN addresses workload
imbalances issues through multiple rounds of runtime load-rebalancing, but
this leads to high inter-PE communication.  Finally, SpMM-based approaches face
more severe load imbalances for implementing GNNs that involve additional
complex computations before Aggregation (e.g., softmax in GATs and DiffPool).
In fact, AWB-GCN targets only GCNs and not general GNNs. 

\noindent
{\bf Novelty of this work:}
We propose the GNNIE (pronounced ``genie'') architecture that uses a {\em
single engine} that efficiently performs both Weighting and Aggregation.  The
GNNIE framework handles high levels of sparsity in the input vertex feature
vectors and the adjacency matrix, with novel approaches for {\em load
balancing} and {\em graph-specific caching}.  It covers a {\em number of GNN
topologies}, from lower accuracy/lower computation (e.g., GCN, GraphSAGE)
higher accuracy/higher computation (e.g., GATs), as motivated in
Fig.~\ref{fig:GNN accuracy comparison}, and is {\em more versatile} than
previous methods in handling functions such as softmax over a neighborhood
(e.g., as used for attention normalization in GATs; prior
work~\cite{autenhardware} on GATs, skips this crucial step).

Novel methods to mitigate sparsity effects, and overcome load imbalances and compute
bottlenecks, include:

\begin{itemize}[leftmargin=*]

\item {\bf \em Load balancing during Weighting} based on splitting vertex
features into blocks (Section~\ref{sec:CPE_scheduling}). Together with load
balancing (Section~\ref{sec:Loadbalancing_weighting}), this enhances
throughput during Weighting by ensuring high PE utilization and skipping
unnecessary computations, by (a)~Reordering computations on a {\em flexible MAC
(FM) architecture} to address imbalances due to input feature vector sparsity
variations. Computations are dynamically mapped to heterogeneous PEs, each with
different numbers of MAC units.  (b)~{\em Static load redistribution} to nearby
PEs, offloading computations from heavily-loaded to lightly-loaded rows,
minimizing inter-PE communication.


\item {\bf \em Load-balanced edge Aggregation}
(Section~\ref{sec:Aggregation_computations}) through a mapping scheme that fully
utilizes the PEs. For GATs, we further propose a novel linear-complexity
computation that implements compute-bound attention vector multiplication
similarly as Weighting, and memory-bound attention coefficient computation to
maximize reuse of cached data.

\item {\bf \em Lightweight graph-specific dynamic caching}
(Section~\ref{sec:Caching}), fetching vertices in unprocessed degree order;
aggregation operates on dynamic subgraphs formed by cached vertices.  This
new lightweight scheme is effective in avoiding the random DRAM accesses that
plague graph computation.

\end{itemize}

\noindent {\bf Speedups:} 
On five GNN datasets, results based on an RTL implementation and cycle-accurate simulation show that even
including all of its preprocessing overheads, GNNIE delivers average speedups of
21233$\times$ over CPUs (Intel Xeon Gold 6132 + PyTorch Geometric), 699$\times$ over
GPUs (V100 Tesla V100S-PCI + PyTorch Geometric) and 35$\times$ over
prior work.

\vspace*{-2mm}
\section{Background}
\label{sec:BckGNN}


\noindent
In layer $l$ of a GNN, each vertex $i$ in the graph is represented by an
$F^l$-dimensional row vector, $\mathbf h_i^l$, called the {\em vertex feature
vector}; $\mathbf h_i^0$ is the input vertex feature vector.
For each vertex $i$ in a layer, over a set of neighboring vertices $j$,
the GNN aggregates information from vectors $\mathbf h_j^{l-1}$ of the
previous layer, and processes it to create the output feature
vector, $\mathbf h_i^l$.

Table~\ref{tbl:allGNNs} shows the Weighting and Aggregation operations for
graph convolution networks (GCNs)~\cite{kipf2016semi},
GraphSAGE~\cite{hamilton2017inductive}, graph attention networks
(GATs)~\cite{velivckovic2017graph_2}, and GINConv~\cite{xu2019ginconv}.

\begin{table}[thb]
\vspace{-2mm}
\centering
\caption{Summary of operations in layer $l$ of various GNNs.}
\vspace{-2mm}
\resizebox{0.92\linewidth}{!}{%
\begin{tabular}{|l|l|}
\hline
GCN &
$\mathbf h_i^l = \sigma \left (
      \textstyle \sum_{j \in \{i\} \cup N(i)}
         \frac{1}{\sqrt{d_i d_j}} \mathbf h_j^{l-1} W^l
      \right )$
\\ \hline
GraphSAGE &
$\mathbf h_i^l = \sigma \left (
        a_{k} \left ( \mathbf h_j^{l-1} W^l 
                              \forall \, j \in \{i\} \cup S_{N(i)}
              \right )
        \right )$
\\ \hline
GAT &
$\mathbf{h}_{i}^{l} = \sigma \left ( 
      \frac{\sum_{j \in \{i\} \cup N(i)} 
           \mbox{exp}(e_{ij}) \mathbf{h}_{j}^{l-1} W^l}
           {\sum_{j \in \{i\} \cup N(i)} \mbox{exp}(e_{ij})}
                            \right )$ \\
& 
$e_{ij} = \mbox{LeakyReLU}(\mathbf a^T \cdot [\mathbf h_i^{l-1} W^l
                      || \mathbf h_j^{l-1} W^l])$
\\ \hline
GINConv &
$\mathbf h_i^l = \mbox{MLP}^l \left ( (1 + \epsilon^l) \mathbf h_i^{l-1} +
                         \textstyle \sum_{j \in {\cal N}(i)} \mathbf h_j^{l-1},
                         W^l, \mathbf b^l
                             \right )$
\\ \hline
\end{tabular}
}
\label{tbl:allGNNs}
\vspace{-2mm}
\end{table}

\noindent
{\bf Weighting} multiplies the vertex feature vector, $\mathbf h_i^{l-1}$ of 
each vertex by a weight matrix, $W^l$, of dimension $F^{l-1} \times F^l$.


\noindent
{\bf Aggregation} combines the weighted vertex feature vectors 
neighboring vertex $i$.  If $N(i)$ is the immediate one-hop neighborhood of
vertex $i$, then for GCNs, GATs, and GINConv, ${\cal N}(i)=\{i\} \cup N(i)$. For
GraphSAGE, ${\cal N}(i) = \{i\} \cup S_{N(i)}$, where $S_{N(i)}$ is a random
sample of $N(i)$. At vertex $i$:

\noindent
\underline{GCNs}: Each product $\mathbf h_j^{l-1} W^l$, $j \in {\cal N}(i)$, is
multiplied by $1/\sqrt{d_i d_j}$ ($d_*$ is the vertex degree).  The result is
summed.

\noindent
\underline{GraphSAGE}: 
The products $\mathbf h_j^{l-1} W^l$ are combined over $j \in {\cal N}(i)$
using aggregator $a_k$ (typically, mean or pooling).

\noindent
\underline{GATs}: 
For each edge $(i,j)$, an inner product with a learned attention vector
$\mathbf a^l$ finds the normalized attention coefficient
\begin{equation*}
\alpha_{ij} = \mbox{softmax} ( \mbox{LeakyReLU}
       ({{\mathbf a}^l}^T \cdot [ 
               \mathbf h_i^{l-1} W^l] || \mathbf h_j^{l-1} W^l] ))
\label{eq:attention_coeff}
\end{equation*}
followed by 
$\textstyle \sum_{j \in \{i\} \cup {\cal N}(i)} e_{ij} \mathbf h_j^{l-1} W^l$,
a weighted aggregation.

\noindent
\underline{GINConv}: The vertex feature vertices of all neighbors of a vertex $i$
are summed and added to $\epsilon^l$ times the vertex feature vector of $i$,
where $\epsilon^l$ is a learned parameter, using a multilayer perceptron (MLP)
with weights $W^l$ and $\mathbf b^l$:
\begin{equation}
\mathbf h_i^l = \mbox{MLP}^l \left ( (1 + \epsilon^l) \mathbf h_i^{l-1} +
                         \textstyle \sum_{j \in {\cal N}(i)} \mathbf h_j^{l-1},
                         W^l, \mathbf b^l
                             \right )
\end{equation}

\noindent
The activation operator $\sigma$ (softmax or ReLU), is applied to the
aggregated weighted vertex feature vector, yielding the updated $\mathbf h_i^l$.
For GINConv, activation is built into the MLP.

GINConv concatenates the sum of all vertex feature vectors across all layers to
obtain a representation for the graph as
\begin{equation}
\mathbf h_G = \Bigg | \Bigg |_{l = 1}^L
         \textstyle \left ( \sum_{i \in G} \mathbf h_i^l \right )
\end{equation}

DiffPool~\cite{ying2018diffpool} can be combined with any of these GNNs to
reduce the volume of data.  It uses two GNNs, one to extract vertex embeddings
for graph classification, and one to extract embeddings for hierarchical
pooling.  The embedding GNN at layer $l$ is a standard GNN with Weighting
and Aggregation,
\begin{equation}
Z^{l-1} = \mbox{GNN}_{embed} (A^{l-1},X^{l-1});
\end{equation}
where $A^{l-1}$ is the adjacency matrix of the coarsened graph at level $(l-1)$,
and $X^{l-1}$ is the matrix of input cluster features.  The pooling GNN
generates the assignment matrix:
\begin{equation}
S^{l-1} = \mbox{softmax} \left ( \mbox{GNN}_{pool} (A^{l-1},X^{l-1}) \right )
\end{equation}
The number of clusters in layer $l$ is fixed during
inference.  The coarsened adjacency matrix $A^l = {S^{(l-1)T}} A^{l-1} {S^{l-1}}$,
and the new embedding matrix $X^l = {S^{(l-1)T}}~Z^{l-1}$.

\vspace{-1mm}
\section{Accelerator Architecture}
\label{sec:HAcc}

\label{sec:archoverview}

\ignore{
\noindent
To design a GNN accelerator, we first analyze the execution of the  Weighting
and Aggregation steps discussed in Section~\ref{sec:BckGNN}.  The operations in
each of these steps are represented by matrix-vector or inner product-vector
multiplications.  Therefore, our hardware platform must be capable of
performing these operations.  
}

\ignore{
Depending on the type of GNN, the accelerator may require some specialized
units such as exponentiators or dividers.  Our architecture is generated using
a parameterizable generator that can create specialized hardware, customized to
a target GNN.  While the most general architecture can implement any GNN, the
efficiency of the platform can be enhanced through this customization, e.g., by
including specialized units only if the target GNN demands them.
}

\begin{figure}[t]
\centering
\vspace{-2mm}
\includegraphics[width=2.8 in]{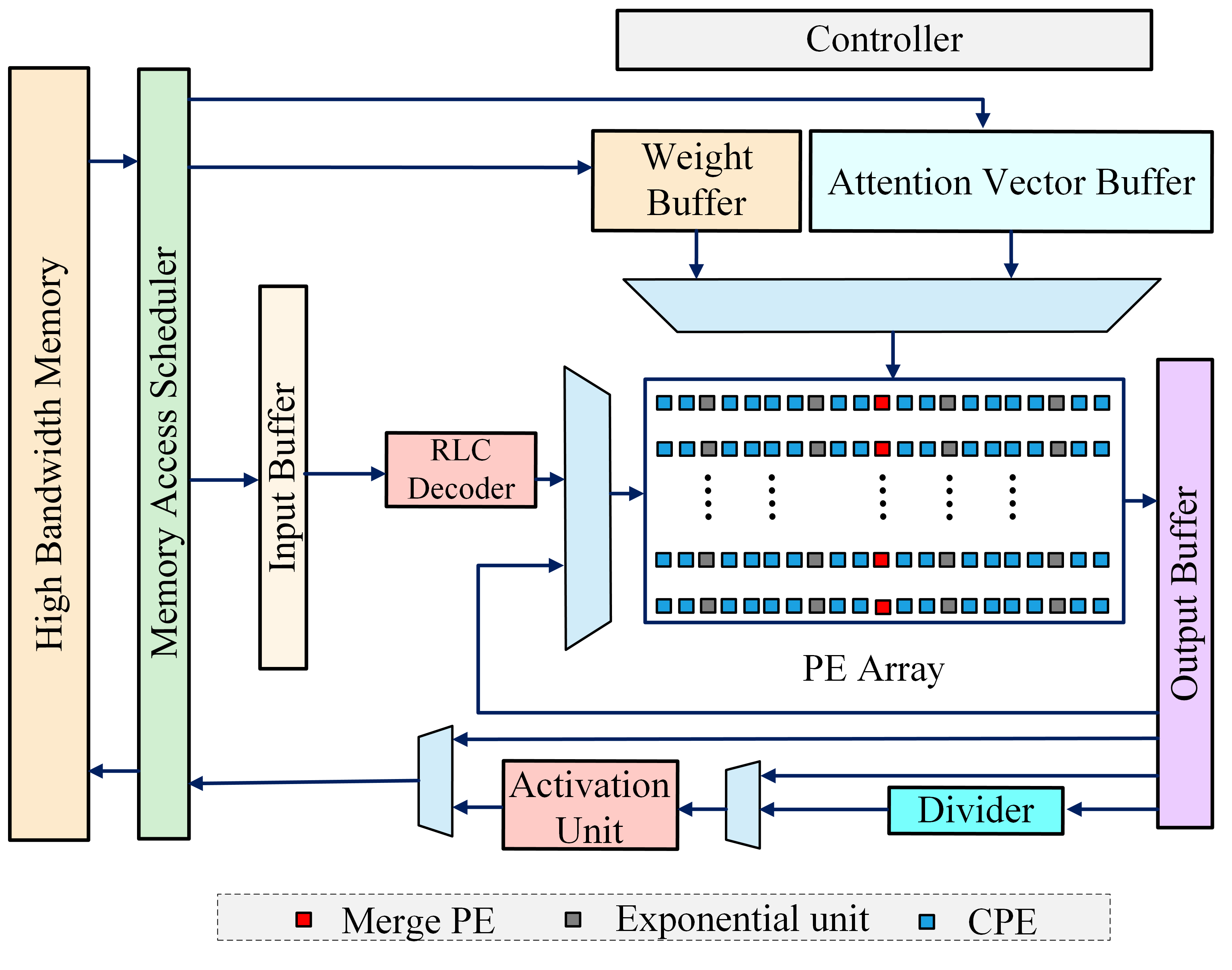}
\vspace{-3mm}
\caption{Block diagram of the proposed architecture.}
\label{fig:block diagram}
\vspace{-6mm}
\end{figure}

\noindent
The block diagram of the proposed accelerator is illustrated in
Fig.~\ref{fig:block diagram}, and it consists of the following key components:

\noindent
(1) {\bf HBM DRAM:} The high-bandwidth memory (HBM) DRAM stores information
about the graph.  The adjacency matrix of the graph represents its connectivity
information and is stored in sparse compressed sparse row (CSR)
format.  Other formats (CISR~\cite{Fowers14}, C$^2$SR~\cite{Srivastava20a},
CISS~\cite{Srivastava20b}) are not viable candidates as they ignore the
underlying graph structure: GNNIE uses adjacency matrix connectivity information
to schedule computations and is not a matrix multiplication method.  

The sparse input vertex feature vectors are encoded using run-length
compression (RLC)~\cite{Salomon07}.  We choose RLC because it is lossless and
the decoder has low power/area overhead: this is important because
it is only used for the input layer and not thereafter.  Alternatives such as
CISS have much higher implementation overhead and have been targeted to
lock-step systolic arrays, which are unsuitable for Weighting due to the 
insertion of stalls to handle feature vector sparsity variations.

The DRAM is also used to store intermediate results that do not fit in on-chip
memory. High bandwidth options such as HBM or GDDR6 are viable
for edge AI~\cite{Hruska19,Ward-Foxton20}.

\noindent
(2) {\bf Memory interface:} 
The {\em input buffer} stores vertex features for one pass of
the current layer $l$, i.e., $\mathbf h_i^{l-1}$ for vertices $i$ being processed,
and the edge connectivity information of the subgraph.  Double-buffering is
used to reduce DRAM latency: off-chip data is fetched while the PE array computes. 

Sparse data is transmitted from off-chip DRAM to the input buffer using 
RLC encoding.  The input buffer keeps this data in RLC format until it is ready
for use, when the data is sent through the {\em RLC decoder} to the PE array.
The RLC decoder is activated for sparse input layer vertex feature vectors,
and bypassed for denser feature vectors in later layers.

The {\em output buffer} caches intermediate results for vertex feature vectors,
including the result of multiplication by $W^l$ after Weighting, and the result
after Aggregation.  The end result is written to off-chip memory.  The
{\em weight buffer} holds the values of the weight matrix $W^l$ during
Weighting, and, for GAT computations, the attention vector during Aggregation.

The {\em memory access scheduler} coordinates off-chip memory requests from the
input/output/weight buffers.

\noindent
(3) {\bf An array of processing elements (PEs):}  The array consists of an $M
\times N$ array of {\em computation PEs} (CPEs). Each CPE has
two scratch pads (spads) and MACs. 

Within the array of CPEs, we merge multiple columns of {\em Special
Function Units (SFUs)} (e.g., exp, leaky ReLU) [grey blocks], and a row of
merge PEs (MPEs) [red blocks].  Interleaved placement allows low latency and
communication overhead with CPEs.  
For exponentiation, we use an accurate, low-area lookup-table-based
implementation~\cite{nilsson2014hardware}.

\ignore{
As stated earlier, the parameterizable array generator
can customize the array to the network: for example, since the exponential
units are used only by the GAT, they may be dropped for a GNN engine that is
not targeted to GATs.
}

{\em Merge PEs (MPEs)} 
are used to aggregate partial results of vertex features sent from the CPE array
during Weighting and Aggregation.  One MPE is dedicated for each CPE column in
the array (Fig.~\ref{fig:block diagram}), collecting partial results from the
CPEs in the column.  Each MPE has one update spad to hold the partial results
sent from the CPEs in its column. The update spad contents are sent to an
accumulator bank. A bank of partial sum (psum) spads holds intermediate results
from the accumulator bank.  When the summation is complete, the psum bank sends
results to the output buffer.  

\noindent
(4) The {\bf Activation unit} performs an activation operation on the vertex features at the
final activation stage of computation.


\noindent
(5) The {\bf controller} coordinates operations, including assigning vertex
features to the CPE, workload reordering among the CPEs, sending CPE results to
the MPEs, sending MPE data to the output buffer, and writing concatenated MPE
data.

For a GCN, the layer-wise computation can be written as:
\begin{equation}
\mathbf h_{i}^l = \sigma ( \widetilde{A} \mathbf h_{i}^{l-1} W^l )
\label{eq:GCN_eq_2}
\end{equation}
Here, $\widetilde{A}=D^{-1/2} (A+I) D^{-1/2}$ is the normalized adjacency
matrix, $I$ is the identity matrix, and $D_{ii}=\sum A_{ij}$. This
can be computed either as
$(\widetilde{A} \times  \mathbf h_{i}^{l-1}) \times W^l$ or
$ \widetilde{A} \times (\mathbf h_{i}^{l-1}  \times W^l)$.
The latter requires an order of magnitude fewer computations than the
former~\cite{Geng2020AWBGCNAG,liang2020engn}, and we use this approach.
Moreover, as $\widetilde{A}$ is highly sparse and shows power-law
behavior, we will perform edge-based Aggregation with optimized graph-specific
cache replacement policies to limit off-chip accesses.

\ignore{
Computations in other GNNs also follow a similar pattern, aggregating
$\mathbf h_{i}^{l-1}\times W^l$, where the Aggregation operation is represented
by $\widetilde{A}$. We use the same strategy.
}

\ignore{
\begin{figure}[t]
\centering
\includegraphics[width=1.7in]{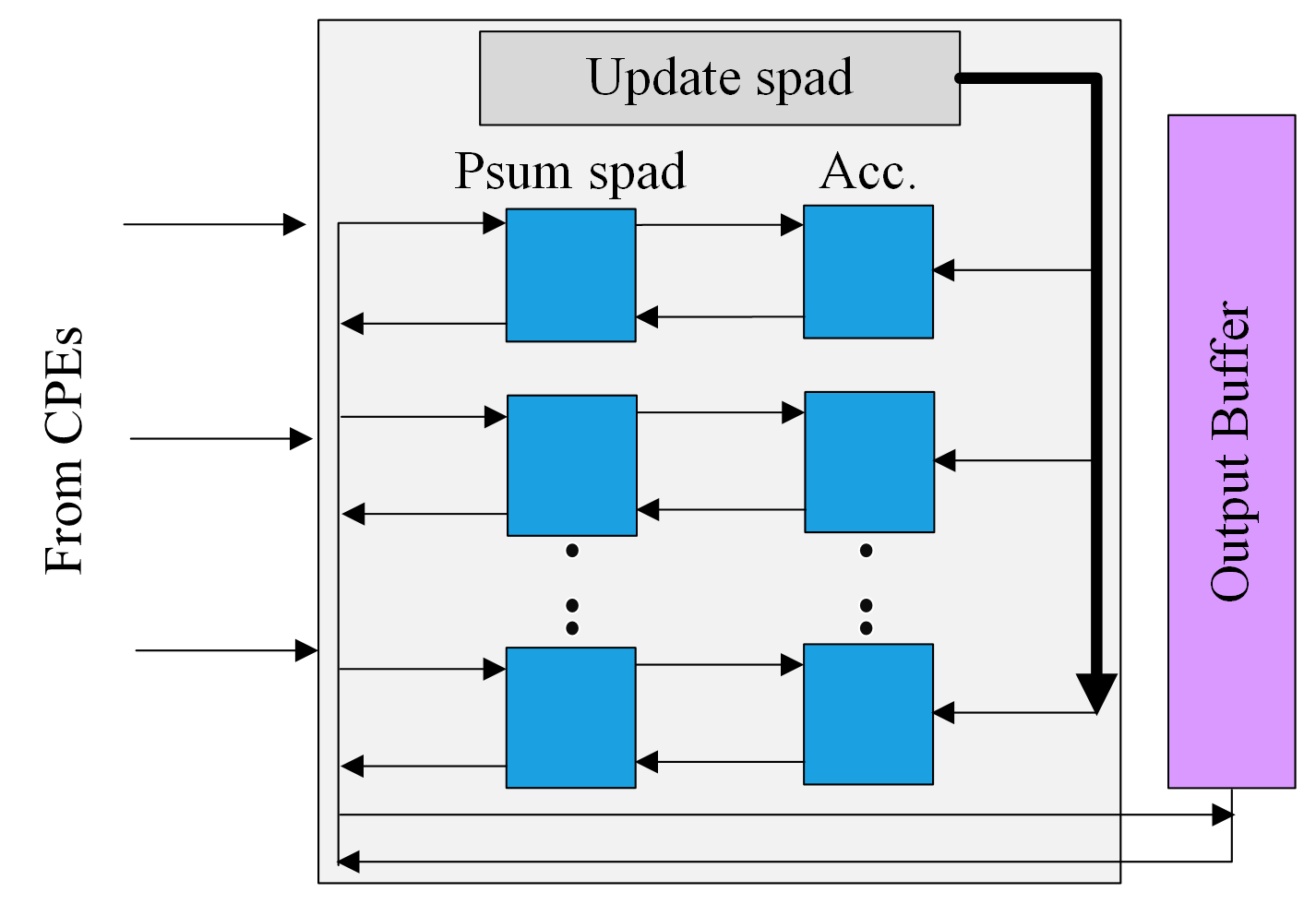}
\caption{Structure of a merge PE.}
\label{fig:Merge_PE}
\vspace{-6mm}
\end{figure}
}

\vspace{-2mm}
\section{Mapping Weighting to CPE{\rm \bf s}}
\label{sec:Mapping}

\vspace{-1mm}
\subsection{Scheduling Operations in the CPEs}
\label{sec:CPE_scheduling}

\noindent
We now map the Weighting step, which multiplies the {\em sparse} feature row vector $\mathbf
h_i^{l-1}$ with the {\em dense} weight matrix $W^l$, to the architecture.  The
feature vectors are fetched from DRAM, core computations are performed in
the CPEs, and the results from the CPEs are assimilated in the MPEs before
being written back to DRAM.  Our novel scheduling methodology keeps the CPEs
busy during the computation, so that Weighting is not memory-bounded. We
partition data in two ways (Fig.~\ref{fig:Weighting}):

\noindent
(1) {\bf Across the vertex feature vector:} 
We process a \blueHL{\bf block} of $k$ elements of $\mathbf h_i^{l-1}$ at a time, and
multiplying it by the corresponding $k$ rows of $W^l$. 
This is mapped to a row of the CPE array.  With a block size
of $k = \left \lceil F^{l-1}/M \right \rceil$, the entire feature vector is
processed in the CPE array.

\noindent
(2) {\bf Across vertices:}
We process feature vectors for a \blueHL{\bf set} of $s$ vertices at a time in the
PE array, as shown in Fig.~\ref{fig:Weighting}, where $s$ is constrained by the
size of the input buffer. To process all vertices in the graph $G(V,E)$, we process
$\lceil |V|/s \rceil$ sets as:
\vspace*{-2mm}
\begin{align}
\hspace*{-2mm}
\mathbf h_i^{l-1} W^l &= \left [ 
        \sum_{i=0}^{F^l-1} \mathbf h_{(0:k-1)}^{l-1}  W_{(0:k-1,i)}^l ,
        \sum_{i=0}^{F^l-1} \mathbf h_{(k:2k-1)}^{l-1} W_{(k:2k-1,i)}^l ,
        \right . \nonumber\\
& \hspace*{0.25in} \left . \cdots , 
        \sum_{i=0}^{F^l-1} \mathbf h_{((N-1)k:F^{l-1})}^{l-1} W_{((N-1)k:F^{l-1}),i}^l
            \right ] 
\label{eq:mvm_2}
\end{align}

\vspace*{-1mm}
\noindent
where the term in each sum is processed in a separate CPE.

\begin{figure}[b]
\vspace{-4mm}
\centering
\includegraphics[width=2.8in]{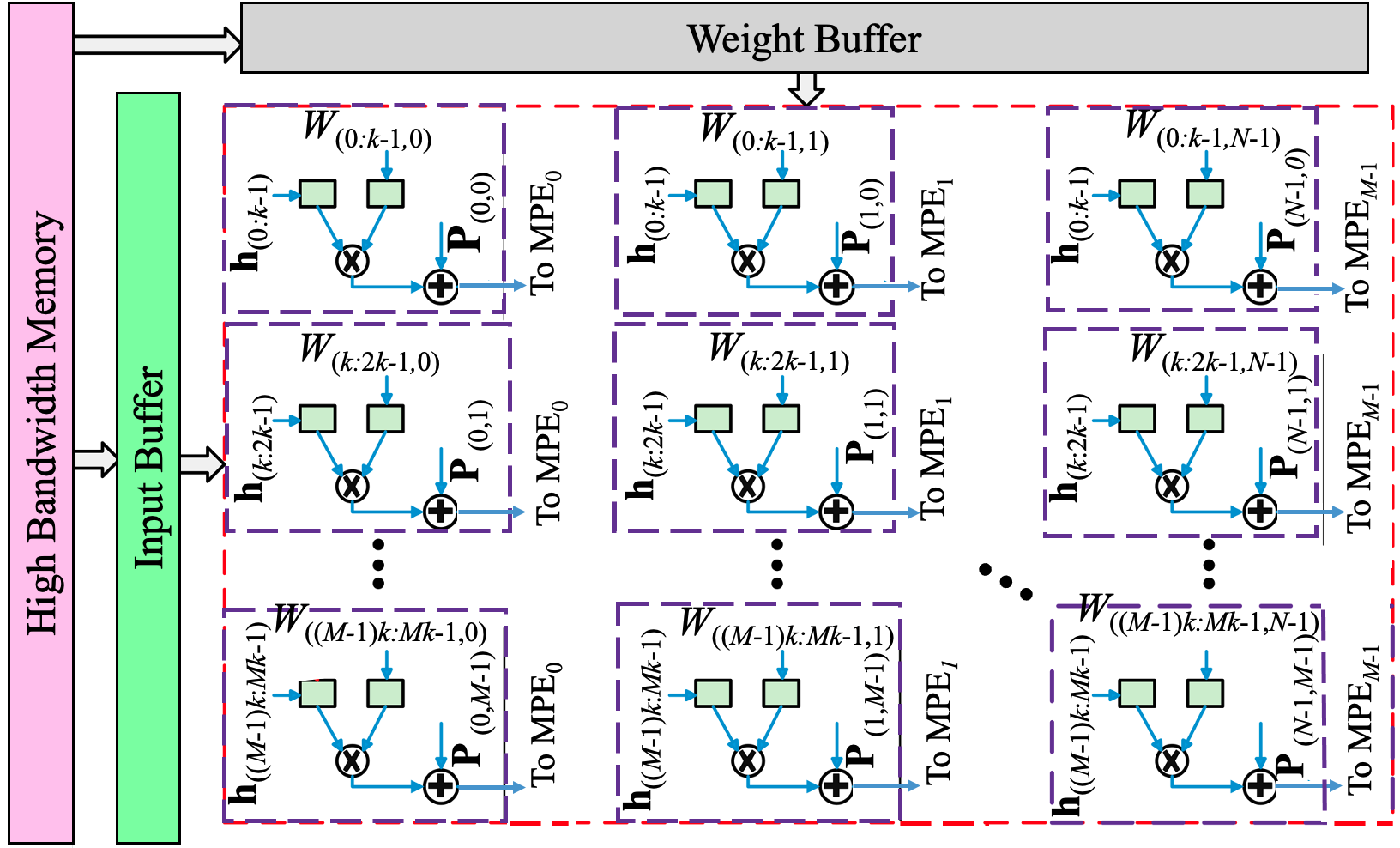}
\vspace{-2mm}
\caption{Weight-stationary linear transformation of vertex features.}
\label{fig:Weight-stationary-data-flow}
\end{figure}

We use a weight-stationary scheme (Fig.~\ref{fig:Weight-stationary-data-flow}).
Each vertex goes through Weighting set by set, placing $k$-element blocks of
the vertex feature vectors for each set into the input buffer.


\begin{figure}[t]
\centering
\hspace*{-0.025\linewidth}
\includegraphics[width=0.90\linewidth]{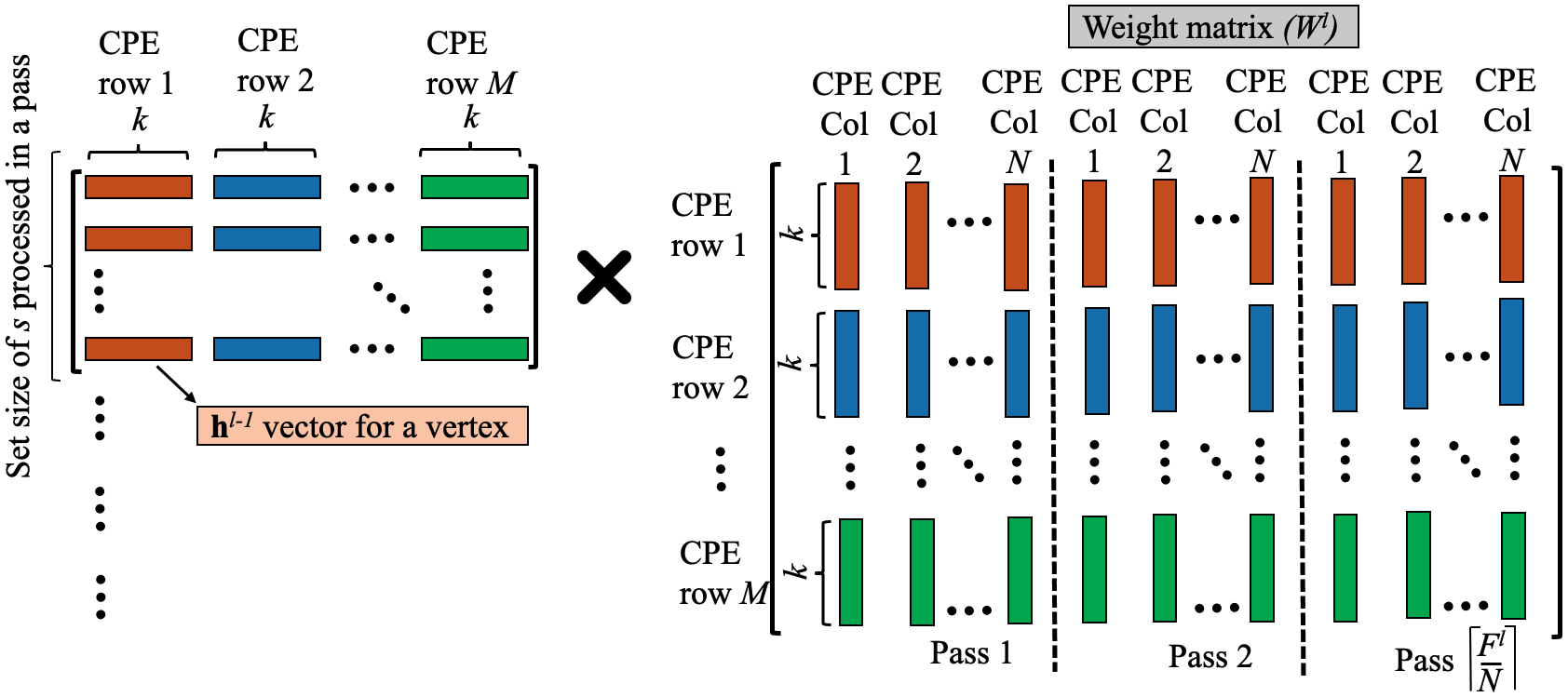}
\vspace{-2mm}
\caption{Mapping Weighting operations to the CPE array.}
\label{fig:Weighting}
\vspace{-6mm}
\end{figure}

We fetch $N$ columns of the weight matrix, $W^l$, from the DRAM to the weight buffer.  A
\blueHL{\bf pass} processes all vertex feature vectors (i.e., processing all
vertices in all sets).  As shown in Fig.~\ref{fig:Weighting}, we multiply the vertex
feature vectors in all sets with $N$ columns of $W^l$ in the pass.
At the end of a pass, the next set of $N$ columns of $W^l$ is
loaded.  After all passes are completed, the current
set of weights is replaced by a new set, and the process continues under the
weight-stationary scheme.  Within each pass, the CPEs are loaded as follows:

\noindent
\vspace{-5mm}
\begin{itemize}[leftmargin=*]
\item
Each column of $W^l$ is loaded to a CPE column in chunks of
$k$ rows, i.e., $W_{(ik:(i+1)k-1,j)}$ is loaded into CPE $(i,j)$.
\item
For a given set of $s$ vertices, the $i^{\rm th}$ subvectors, of size $k$, of all
$s$ vertex feature vectors are broadcast to the entire CPE row $i$.  This is
indicated by $\mathbf h$ in Fig.~\ref{fig:Weight-stationary-data-flow}.
\end{itemize}
To leverage input data sparsity, a zero detection buffer is used 
to skip CPE computations involving zeros.  
Completed CPE results are sent to the MPE for accumulation, and the
next block of size $k$ 
is loaded to the CPE.

\noindent
{\bf Benefit of using vertex feature subvector blocks:} Our use of
$k$-element blocks instead of the entire vector allows a CPE to skip zero
subvectors during pipelined execution and immediately move on to a block from
the next available subvector.  The next block will be fetched from the input
buffer, and under the weight-stationary scheme, it can start computation with
the already-loaded weights in the CPE.  

The proposed weight-stationary dataflow maximizes the reuse of the weights
cached in the weight buffer, which in turn reduces the size requirement of the
on-chip weight buffer. Though the feature vectors fetched in the input buffer
are get reused, for all datasets evaluated, the computation time for vertices
cached in the input buffer is seen to be larger than the memory fetch time
under the HBM 2.0 off-chip bandwidth.

\vspace{-2mm}
\subsection{MPE Processing and Weight Updates}

\noindent
The MAC operation within each CPE generates a partial result for an element of the
transformed vertex features.  This is sent to the MPE in its column for
accumulation over the vertex feature subvectors, along with a tag that denotes
its vertex.  Due to the irregular completion times for the CPEs, the MPE may
accumulate partial sums for several vertices at a time.  A bank of psum buffers
holds the partially accumulated results: when all partial sums are accumulated
for a vertex feature vector, the MPE sends the result to the output buffer,
along with the vertex ID $i$: this is one element of the result of multiplying
the feature vector of vertex $i$ and $W^l$.  When all $F^l$ elements are
computed, the result is written back to DRAM.

After a CPE column processes all feature blocks for all vertices, the next pass
begins. The weights in that column are replaced with the next column of weights
from $W^l$. To overlap computations and keep the CPEs busy, we use
double-buffering to fetch the next block of weights from the DRAM to the chip
while the CPEs perform their computations.  

\vspace{-2mm}
\subsection{Load Balancing for Weighting}
\label{sec:Loadbalancing_weighting}

\noindent
The Weighting computation skips zeros in the vertex feature vector.  Vertex feature
vectors in the input layer have different sparsity levels (e.g., in Regions A and B of
Fig.~\ref{fig:input feature vector sparsity}), and this is also true of the
$k$-subvectors.  Hence, some $k$-subvectors are processed rapidly (``rabbits'')
while others take longer (``turtles'').  This causes workload imbalance in the
CPE array.

The MPEs that accumulate the results of the CPEs must keep track of psums from
a large number of vertices, but they have only limited psum slots for accumulating
information.  The rabbit/turtle disparity implies that stalls may have to be
introduced to stay within the limits of available psum memory in the MPE.  As
results are accumulated in the output buffer, a larger number of vertex feature
vectors must be stored within the buffer, waiting to be completed and written
to the DRAM, to account for the disparity between rabbits and turtles.

\noindent
{\bf Flexible MAC (FM) Architecture:}
We can avoid stalls and speed up computation with more MACs per CPE.
Increasing the number of MACs per CPE uniformly throughout the array overcomes
the bottleneck of ``turtles,'' but is overkill for ``rabbits.'' Our flexible
MAC architecture uses a heterogeneous number of MAC units per CPE in different
rows of the array.  The CPE array is divided into $g$ row groups, each with an
equal number of rows; the number of MACs per CPE, $|MAC|_i$, is monotonically
nondecreasing from the first row to the last, i.e., $|MAC|_1 \leq |MAC|_2 \leq
\cdots \leq |MAC|_g$.

The input buffer has a scheduler that assigns vertex feature vectors to CPE
rows.  The scheduler uses information about the total nonzero workload for
each $k$-element block of the vertex feature vector to assign the workload to
CPE rows. The workloads for the $k$-element blocks are first binned based on
the number of nonzeros, where the number of bins equals the number of CPE
groups.  Workload binning is carried out as a preprocessing
step in linear time.
The bin with fewest nonzeros is sent to the first CPE group with fewest MACs,
and so on; the bin with the most nonzeros is sent to the last CPE row group
with the most MACs.  

\begin{figure}[t] \centering
\includegraphics[width=0.8\linewidth]{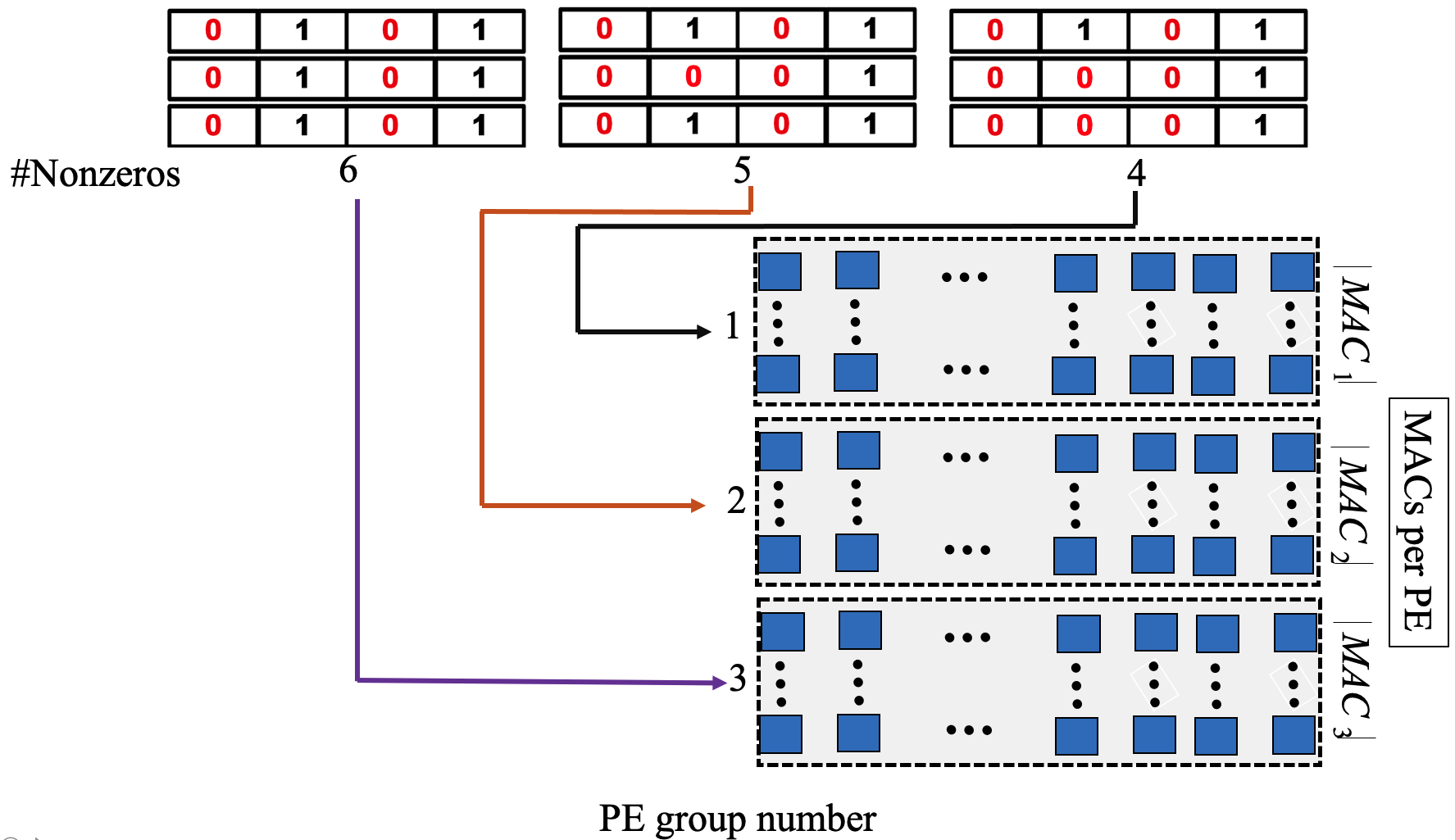}
\vspace{-2mm}
\caption{Workload reordering in flexible MAC (FM) approach.}
\label{fig:workload_reorder}
\vspace{-6mm}
\end{figure}
   
An example of workload reordering among CPE rows is shown in
Fig.~\ref{fig:workload_reorder}.  The CPE array is divided into three groups,
Group 1, 2, and 3, where Group $i$ is equipped with $|MAC|_i$ MACs per CPE,
where $|MAC|_1 < |MAC|_2 < |MAC|_3$.  The vertex feature blocks are binned into
three bins that will be assigned to each group.  Each bin has several vertex
feature blocks: the vertex feature blocks in the left-most bin have the most
nonzeros (six), and those of the right-most bin have the fewest of nonzeros
(four). We see that the least populated bin is assigned to
the group with the fewest MACs, the next to the group with the next number of
MACs, and so on.

\noindent
{\bf Load Redistribution (LR):}
The FM approach does not completely balance the workload.  For greater
uniformity, we redistribute loads among nearby CPEs.  Based on workload
distribution in CPE rows, the controller selects pairs of CPE rows to perform
workload redistribution, offloading a portion of workload from heavily loaded
to lightly loaded CPE rows. 

To perform computation on the offloaded workloads, the weights 
must be transferred with the data.  To minimize communication overhead,
we first finish the computation in FM, to the point where the current weights
are no longer needed, before applying LR. The spads for weights in these CPE
rows are loaded with weights for the offloaded workloads. 



\vspace{-1mm}
\section{Aggregation Computations}
\label{sec:Aggregation_computations}

\noindent
For most GNNs in Section~\ref{sec:BckGNN}, Aggregation is a simple summation
over the neighbors of the vertex, but GATs require significantly more
computation in determining attention coefficients, which are used for weighted
aggregation.  The first two subsections focus on GAT-specific computations. We
then consider Aggregation operations that affect all GNNs.

\vspace{-2mm}
\subsection{Reordering for Linear Computational Complexity}
\label{sec:reorder_alphaij}

\noindent
We present a new method for reordering GAT computations for efficient hardware
implementation.  We define the weighted vertex attention vector for vertex $p$ as
$\boldsymbol{\eta_w}_p^l = \mathbf h_p^{l-1} W^l$.  The first step in finding
the attention coefficient $\alpha_{ij}$ for neighboring vertices $i$ and $j$,
is to multiply the $2F^l$-dimensional attention vector, $\mathbf a^l$, by a
concatenation of two $F^l$-dimensional weighted vertex feature vectors,
($\boldsymbol{\eta_w}_i^l$, $\boldsymbol{\eta_w}_j^l$).  We now show how this
operation is carried out in the PE array by the CPEs.  

Rewriting $\mathbf a^l = [\mathbf a_1^l \; \mathbf a_2^l]$, where $\mathbf
a_q^l$ is the subvector that multiplies $\boldsymbol{\eta_w}_q^l$, we can denote
this inner product as

\vspace{-4mm}
\begin{equation}
e_{ij} 
= {\mathbf a}_1^{l\;T} \cdot \boldsymbol{\eta_w}_i^l +
		{\mathbf a}_2^{l\;T} \cdot \boldsymbol{\eta_w}_j^l
= e_{i,1} + e_{j,2}
\label{eq:attention_coeff_calc}
\end{equation}

\vspace{-1mm}
\noindent
where $e_{i,1} = {\mathbf a}_1^{l\;T} \cdot \boldsymbol{\eta_w}_i^l$, $e_{j,2}
= {\mathbf a}_2^{l\;T} \cdot \boldsymbol{\eta_w}_j^l$,
This goes through a LeakyReLU and then a softmax over all neighbors of $i$ to
find the normalized attention coefficient,

\vspace{-6mm}
\begin{eqnarray}
\alpha_{ij} &=& \mbox{softmax} \left ( \mbox{LeakyReLU} (e_{ij}) \right )
\label{eq:attention_coeff_reform}
\end{eqnarray}

\vspace{-2mm}
\noindent
A na\"ive approach would fetch $\boldsymbol{\eta_w}_j^l$ from each neighbor $j$
of $i$, compute $e_{ij}$ using~\eqref{eq:attention_coeff_calc}, and perform
softmax to find $\alpha_{ij}$.  
%
%
However, since $e_{j,2}$ is required by every vertex for which $j$ is a
neighbor (not just $i$), this would needlessly recompute its value at each
neighbor of $j$.  To avoid redundant calculations, we reorder the computation:
for each vertex $i$, we compute\\
(a)~$e_{i,1} = \mathbf a_1^{l\;T} \boldsymbol{\eta_w}_i^l$, used
to compute $\alpha_{i*}$ at vertex $i$.\\
(b)~$e_{i,2} =  \mathbf a_2^{l\;T} \boldsymbol{\eta_w}_i^l$, used
by all vertices $j$ for which $i$ is a neighbor, to compute $\alpha_{j*}$ at vertex $j$.\\
Since $\mathbf a^l = [\mathbf a_1^l \; \mathbf a_2^l]$ is identical for each
vertex, we calculate $e_{i,2}$ just once at vertex $i$, and transmit it to
vertices $j$.

For $|V|$ vertices and $|E|$ edges, the na\"ive computation performs $O(|E|)$
multiplications and memory accesses to $\boldsymbol{\eta_w}_i^l$) per vertex,
for a total cost of $O(|V||E|)$.  Our reordered computation is $O(|V|+|E|)$,
with $O(|E|)$ accumulations over all vertices, i.e., \underline{\em latency and power are
linear in graph size.}

\vspace{-2mm}
\subsection{Mapping Attention Vector Multiplication}
\label{sec:map_vector_alphaij}

\noindent
As in Weighting, we use a block strategy to distribute computation
in the CPE array.  The vector $\boldsymbol{\eta_w}_i$ is distributed across all $N$
columns of a row, so that the size of each block allocated to a CPE for vertex
$i$ is $G = \lceil F^l/N \rceil$.  Each CPE column processes $V_a$ vertices. 
Here, $V_a$ depends on the number of columns $N$ in the CPE array,
and also depends on the size of the output buffer $|OB|$, i.e., the size of the
set of vertices that can be cached in the output buffer: $V_a = |OB|/N$.

This dot product computation is very similar to the weight-stationary scheme
used in the Weighting step, i.e., the attention vectors remain stationary until
a pass through all the vertices.  The $F^l$-dimensional subvector $\mathbf a_1^l$
is divided into $N$ blocks of size $G$ and distributed columnwise to one of the
spads in each CPE. Vertex feature blocks for $V_a$ vertices at a time, divided
into chunks of size $G$, are loaded into the other spad, and the inner product
computation proceeds.  Since $\boldsymbol{\eta_w}_j^l$ and $\mathbf{a}^l$ are
dense, load balancing in the CPE array is unnecessary.

\ignore{
\begin{figure}[t]
\centering
\includegraphics[width=0.9\linewidth]{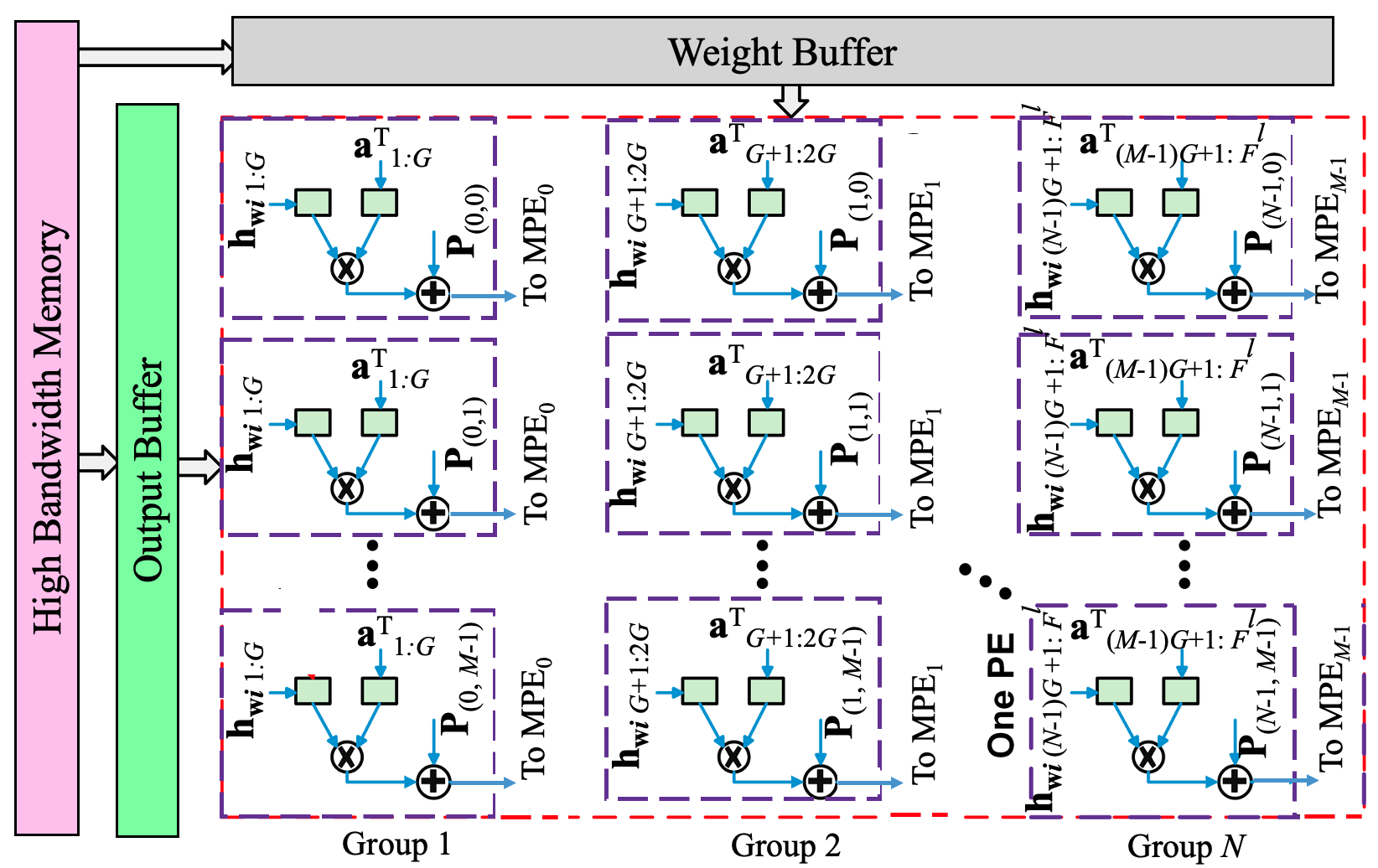}
\caption{Data flow for attention vector multiplication.}
\label{fig:Map_2}
\vspace{-6mm}
\end{figure}
}
As the CPEs in a column finish computation for a vertex, the partial results are sent
to the corresponding MPE for Aggregation. We overlap the computation in a CPE
column and with the Aggregation in the corresponding MPE: as the MPE
aggregates partial results for the current vertex, the blocks of the next
weighted vertex features are loaded into the CPE. Thus, all CPEs and MPEs remain busy.

After all $V_a$ vertices in the row are processed, the spad that contains $\mathbf
a_1^l$ is loaded with $\mathbf a_2^l$, and the second inner product computation
for the $V_a$ vertices is performed, reusing $\boldsymbol{\eta_w}$. The
computed $e_{i,1}$ and $e_{i,2}$ are written back to the output buffer and are
appended to the feature vector of vertex $i$.

\vspace{-2mm}
\subsection{Mapping Edge-based Computations}
\label{sec:map_edge}

\noindent
The last step requires edge aggregation from each neighbor of a vertex.  All
GNNs, perform edge-based summations followed by an activation function; for
GATs, the weights for this summation are computed using methods in the above
subsections.

Typical graphs are too large for the on-chip buffers.  We use a dynamic scheme
(Section \ref{sec:Caching}) to process a subgraph of the graph at a
time, processing edges in parallel in the CPE array.

\noindent
{\bf Load Distribution:} The Aggregation computation brings data into the input
buffer. For each vertex in the subgraph corresponding to the vertices in
the buffer, it accumulates edge data by pairwise assignment to CPE spads.  

Due to power-law behavior, the vertex degrees in the subgraph may have a large
range.  To distribute the load, the Aggregation summations are divided into
unit pairwise summations and assigned to CPEs. For instance, accumulation of a
sum effectively implements an adder tree in which the number of CPEs required
to process Aggregation for each vertex depends on its degree in the subgraph.

\noindent
{\bf GATs:} The final step in computing the attention coefficient $\alpha_{ij}$
involves edge-based computations (Equation~\eqref{eq:attention_coeff_reform}): 

\vspace{-1mm}
\begin{itemize}
\item
the addition, $e_{ij} = e_{i,1} + e_{j,2}$
\item
a LeakyReLU step, LeakyReLU($e_{ij}$)
\item
a softmax step,
$\exp(e_{ij}) \boldsymbol{\eta_w}_j/\sum_{k \in \{i\}\cup N(i)} \exp(e_{ik})$
\end{itemize}
Each edge from a neighbor $j$ to vertex $i$ contributes an $e_{ij}$ to
the numerator of the softmax, and one to the denominator.  These computations
are parallelized in the CPEs among incoming edges of a vertex using
pull-based aggregation~\cite{malicevic2017everything}. 

\begin{figure}[h]
     \centering
     \includegraphics[width=2.6in]{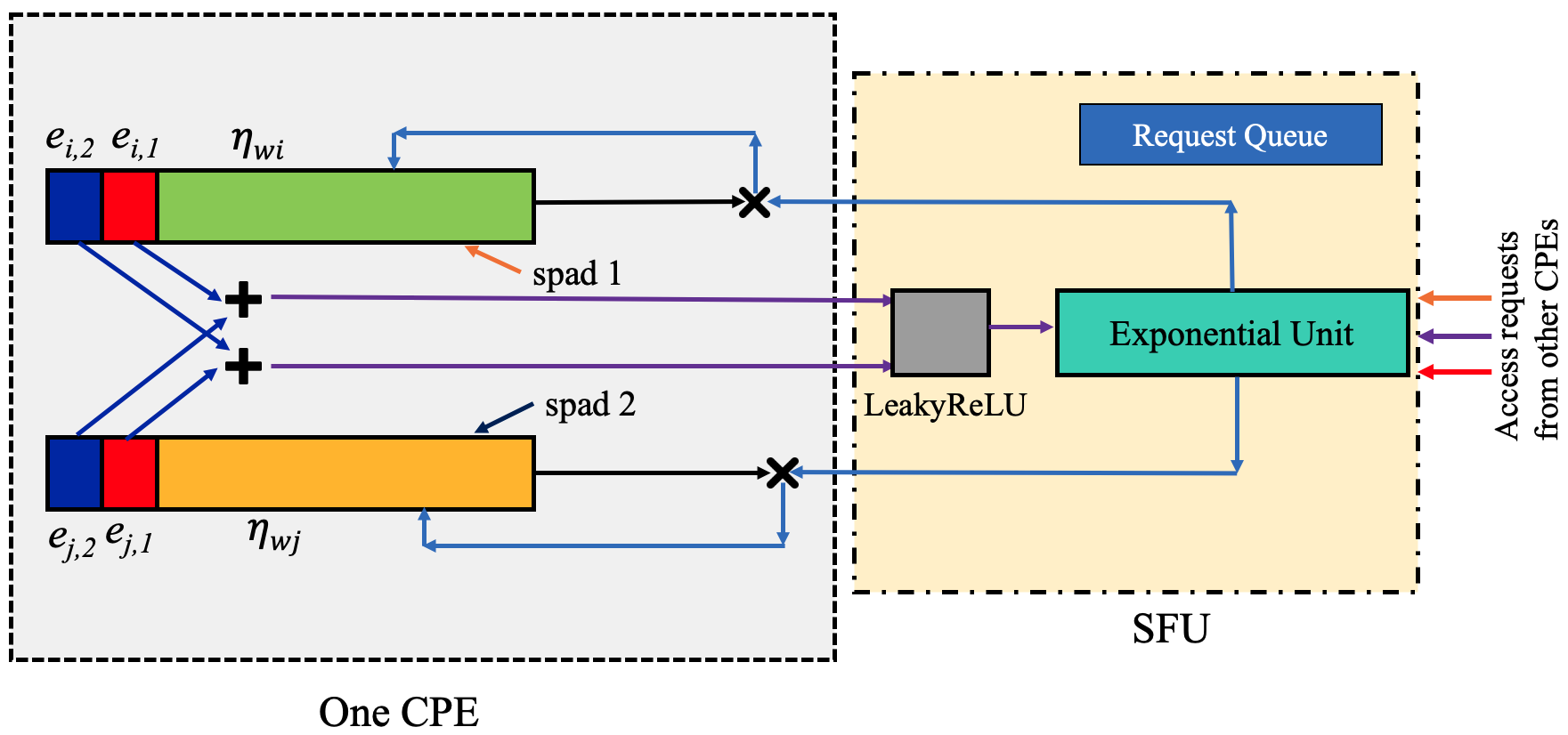}
      \vspace{-1mm}
     \caption{Data flow corresponding to computation of an edge.}
    \label{fig:edge_computation}
\vspace{-6mm}   
\end{figure}

The computation of numerator in the softmax step is shown in
Fig.~\ref{fig:edge_computation}. For a target vertex $i$ connected to a
neighbor $j$ by edge $(i,j)$, $\boldsymbol{\eta_w}_i$, $e_{i,1}$, and
$e_{i,2}$, are loaded into one spad of a CPE, and the corresponding data for
$j$ into the other spad.  For vertex $i$, the result $e_{i,1}+e_{j,2}$ is sent
to the SFU to perform LeakyReLU followed by exponentiation. The output returns
to the CPE and is multiplied with $\boldsymbol{\eta_w}_j^l$. A similar
operation is performed for vertex $j$ to compute $\exp(e_{ji})
\boldsymbol{\eta_w}_i^l$.  
 
\noindent {\bf Other GNNs:} The Aggregation step for GCN, GraphSAGE, GAT and
GINConv involves a sum of weighted vertex feature vectors over all neighbors
$j$ (or a sample of neighbors for GraphSAGE) of each vertex $i$.  This
computation is similar to but simpler than that in
Fig.~\ref{fig:edge_computation}: just addition is performed.

As before, a subgraph of the larger graph is processed at a time.  In
processing vertex $i$, the data for all neighbors $j$ is processed in an adder
tree, placing operands in spad1 and spad2 of a CPE, and storing
the result in spad1.  The partial results for a vertex (partial sum for a
general GNN, or the summed numerator and softmax denominator for a GAT) are
written to the output buffer after each edge computation.  For a GAT, the
values of $\exp(e_{ik})$ are also added over the neighborhood to create the
denominator for the softmax.  Finally, the accumulation over neighbors is
divided by the denominator, in the SFU to obtain the result.
When all components of the sum for vertex $i$ are accumulated, the result is
sent through the Activation unit and written to DRAM.

\vspace{-2mm}
\section{Graph-Specific Caching}
\label{sec:Caching}

\label{sec:caching_aggregation}

\noindent
Aggregation operations intensively access the graph adjacency matrix.
Computational efficiency requires graph-specific caching techniques to transfer
data to/from on-chip input and output buffers, maximizing data reuse and
minimizing off-chip random memory accesses.  A notable feature of our proposed
policy is a guarantee that {\em all random-access patterns are confined to
on-chip buffers} and {\em off-chip fetches are sequential}.

\ignore{
We favor CSR over recent approaches that
optimize memory accesses~\cite{Fowers14,Srivastava20a,Srivastava20b} for SpMM
operations; as we will discuss in Section~\ref{sec:Related}, these are not
optimal for graph data.
}
\ignore{
Our graph-centric approach for Aggregation overcomes the limitations of
HyGCN~\cite{yan2020hygcn}, where the corresponding window sliding/shrinking
technique based on shards has limited parallelism and re-use opportunities, as
outlined in Section~\ref{sec:Intro}.  In comparison with the SpMM-based
AWB-GCN~\cite{Geng2020AWBGCNAG}, where memory accesses are agnostic to the
graph structure, our graph-centric scheme results in fewer accesses to the
adjacency matrix and naturally handles the power-law structure of typical graph
datasets using graph-specific caching.
}

As stated earlier, the adjacency matrix is stored in the CSR format.
Our input is a graph represented by three arrays: (i)~the coordinate array 
lists the incoming/outgoing neighbors of each vertex,
(ii)~the offset array contains the starting offset of each vertex in the
coordinate array, and (iii)~the property array with the weighted vertex feature,
$\boldsymbol{\eta_w}_i^l$ (see Section~\ref{sec:reorder_alphaij}), for
each vertex $i$; for GATs, this is concatenated with $\{e_{i,1},e_{i,2}\}$.

\noindent

\noindent
{\bf Subgraph in the Input Buffer:}
Edge-mapped computations involve a graph traversal to aggregate information
from neighbors.  At any time, a set of vertices resides in the input buffer:
these vertices, and the edges between them, form a subgraph of the original
graph.  In each iteration, we process edges in the subgraph to perform partial
Aggregation operations (Section~\ref{sec:map_edge}) for the vertices in the
subgraph. Under our proposed caching strategy, ultimately all edges in the graph
will be processed, completing Aggregation for all vertices.

\begin{figure}[t]
\centering
\includegraphics[width=0.96\linewidth]{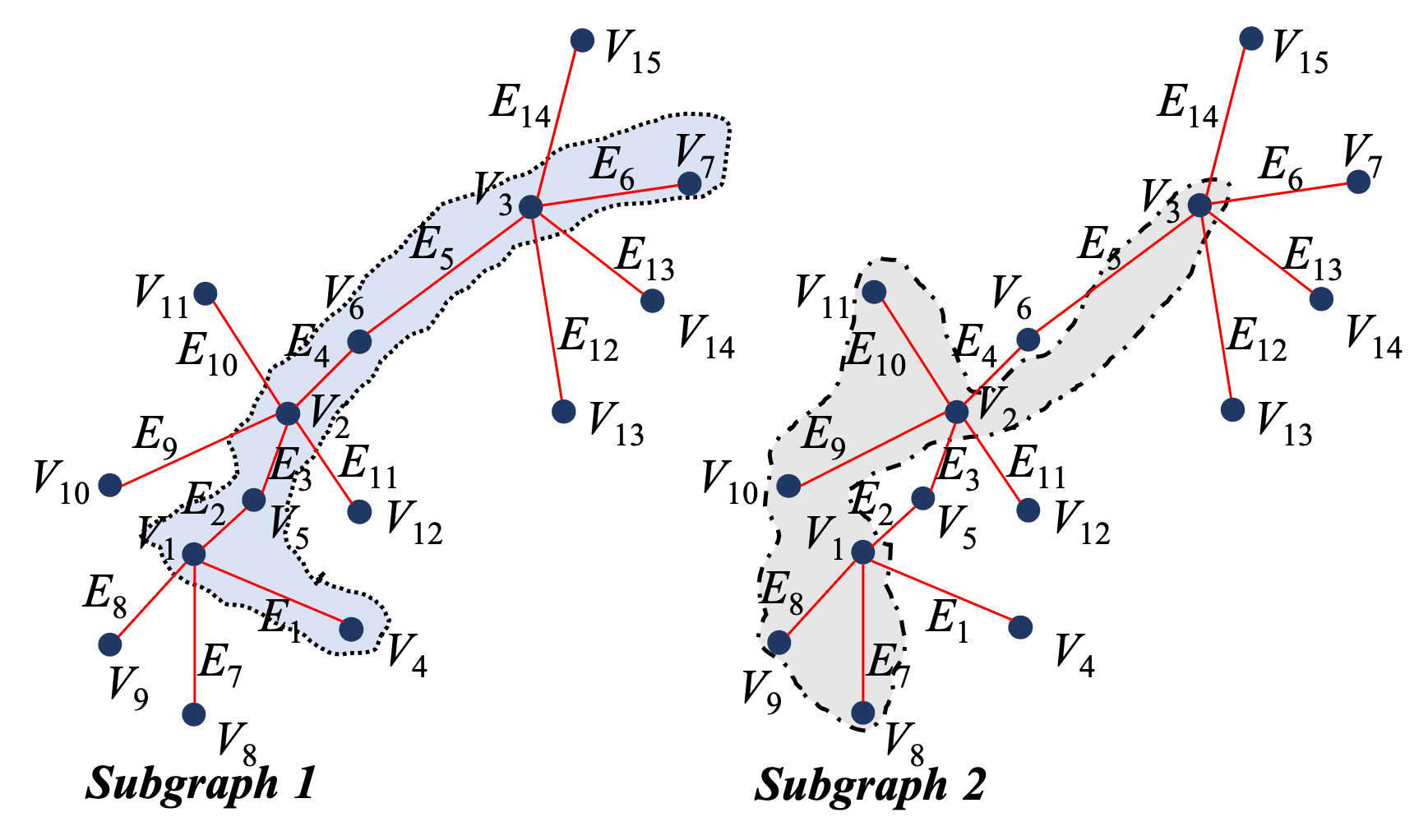}
\vspace{-4mm}
\caption{Example illustrating the subgraph in the input buffer (left) and its
evolution after cache replacement (right).}
\label{fig:PartitionExample}
\vspace{-3mm}
\end{figure}

We illustrate the concept through an example in
Fig.~\ref{fig:PartitionExample}, showing a graph with vertices $V_1$ through
$V_{15}$.  The highest degree vertices are first brought into the cache, i.e.,
the input buffer: vertices $V_1$, $V_2$, and $V_3$ of degree 3, vertices $V_5$
and $V_6$ of degree 2, and then two vertices of degree 1, $V_4$ and $V_7$. The
subgraph, Subgraph~1, consists of these vertices and edges $E_1$ to $E_6$ which
connect them.  After edges $E_1$ through $E_6$ are processed, vertices $V_4$
through $V_7$ have no unprocessed edges and may be replaced in the cache by
$V_8$ through $V_{11}$ in Iteration~2.  This creates Subgraph~2, the subgraph
with edges $E_7$ through $E_{10}$), which is processed next, and so on.

\noindent
{\bf Cache Replacement Policy:}
As vertices are replaced after computation of each subgraph, a replacement
policy is necessary. Our policy prioritizes vertices with the most {\em
unprocessed} edges for retention in the input buffer.  Since such vertices
appear more frequently in the list of neighbors for other vertices in the
coordinate array, this increases the likelihood of finding both the
source and destination of edges in the cache.

The policy requires inexpensive preprocessing to sort vertices 
in order of their degrees.  In practice, it is enough to sort vertices into
bins based to their degrees, differentiating high-degree vertices from
medium-/low-degree vertices to prioritize higher-degree vertices.
After preprocessing, vertices of the input graph are stored
contiguously in DRAM in descending degree order of the bins. 
are broken in dictionary order of vertex IDs.  {\em The key to avoiding
random-access fetches from DRAM is the preprocessing step and the replacement
policy.}

We track the number of unprocessed edges, $\alpha_i$ for vertex $i$,
decrementing it as each neighbor is processed. Initially $\alpha_i$ is the
vertex degree; when $\alpha_i = 0$, ${\bf h}_i^l$ is fully computed.
Tracking $\alpha_i$ requires minimal hardware overhead (a decrementer and one
word of storage per vertex), and its tracking enables GNNIE to maximize edge
processing in each iteration.

\begin{figure}[t]
\centering
\includegraphics[width=\linewidth]{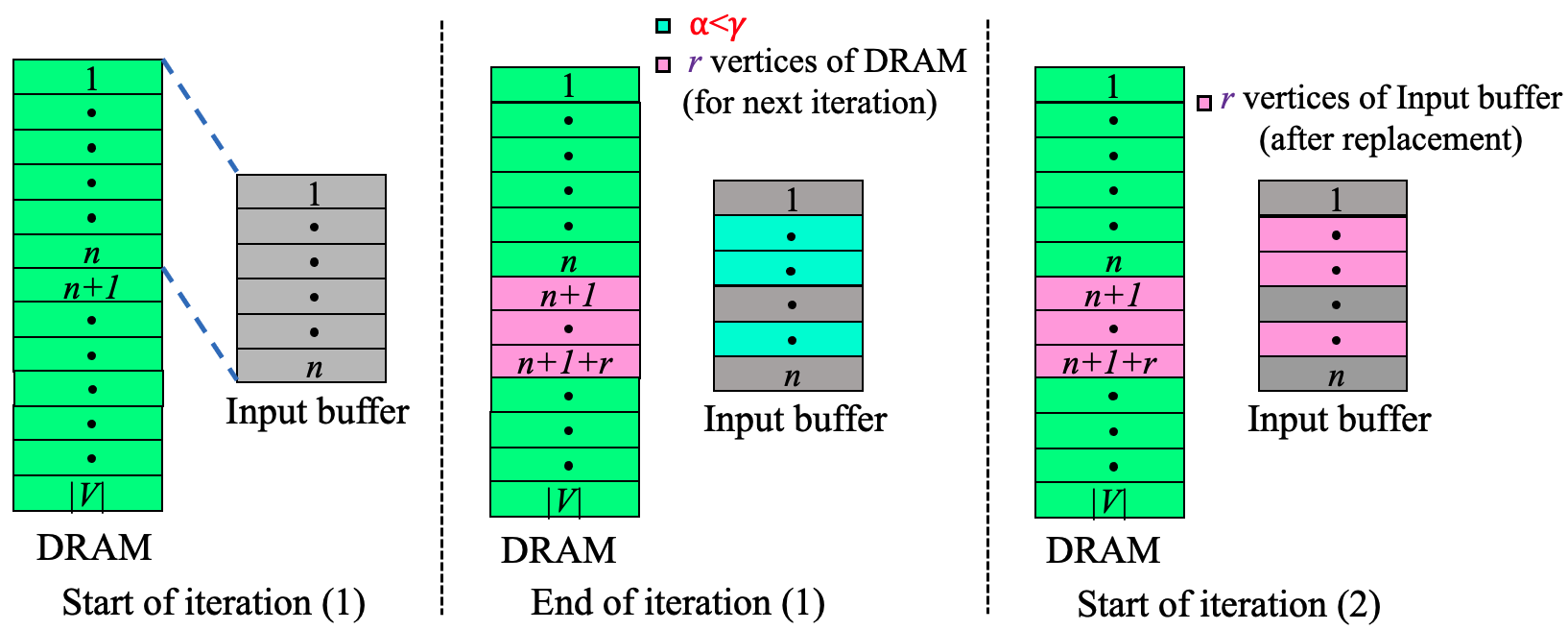}
\caption{Input buffer replacement policy during Aggregation.}
\label{fig:cache replacement policy}
\vspace{-6mm}
\end{figure}

Fig.~\ref{fig:cache replacement policy} illustrates our policy, managed by a
cache controller using a 4-way set associative cache.  Graph vertices
are stored contiguously in DRAM in descending degree order, where
vertex 1 has the highest degree. 
If the input buffer capacity is $n$ vertices, initially data (i.e., feature
vector, connectivity information, $\alpha_i$) for vertices~1 to $n$ are
loaded from DRAM. 

The algorithm processes each such set of vertices in the input buffer in an iteration.
This continues until all vertices are processed.  At the end of iteration~1, i.e.,
after finishing computation of the subgraph corresponding to the first $n$
vertices, if $\alpha_i < \gamma$ for any vertex, where $\gamma$ is a predefined
threshold, it is replaced from the cache.  We replace $r$ vertices in
each iteration using dictionary order if fewer (or more) than $r$ candidates
are available.  These vertices are replaced in the input buffer by vertices $(n + 1)$
to $(n + 1 + r)$ from DRAM: these have the next highest vertex degrees.  For
each such vertex $i$, we write back the $\alpha_i$ value into DRAM.  When all
vertices are processed once, we have completed a {\bf Round}.


Similarly, the partial sums for the vertex feature vector in the output
buffer are updated as more edges in the subgraphs are processed. Any ${\bf h}_i^l$ for
which all accumulations are complete is written back to DRAM. Due to limited
output buffer capacity, and only a subset of partial vertex feature vector sums
can be retained in the buffer, and the rest must be written to off-chip DRAM.
To reduce the cost of off-chip access, we use a degree-based criterion for
prioritizing writes to the output buffer vs. DRAM.  
As partial Aggregation results for softmax are written to DRAM, the numerator
and denominator components for a vertex are stored nearby, for
locality during future fetches.

\noindent
{\bf How our policy avoids random-access DRAM fetches:}
Our policy makes random accesses only to the input buffer; all
DRAM fetches are sequential.  In the first Round, data is fetched from
consecutive DRAM locations.  In the CPE array, aggregation of each vertex
fetches the vertex feature data of its neighbors in the current subgraph in the
cache. Each vertex feature vector may be thus fetched by the CPE array multiple
times according to the graph neighborhood structure, but all such random
accesses are limited to the cache, which has much better random-access
bandwidth than the off-chip memory.

Vertices evicted from the cache, with $\alpha_i < \gamma$, may be fetched
again in a subsequent Round. Even in these Rounds, data blocks are brought into
cache in serial order from DRAM: there are no random accesses from
DRAM.  During DRAM fetches, a cache block is skipped if all of its vertices are
fully processed. The total unprocessed edges in a cache block is tracked
through inexpensive hardware, similar to tracking $\alpha_i$.

\begin{figure}[t]
\centering
\includegraphics[width=0.7\linewidth]{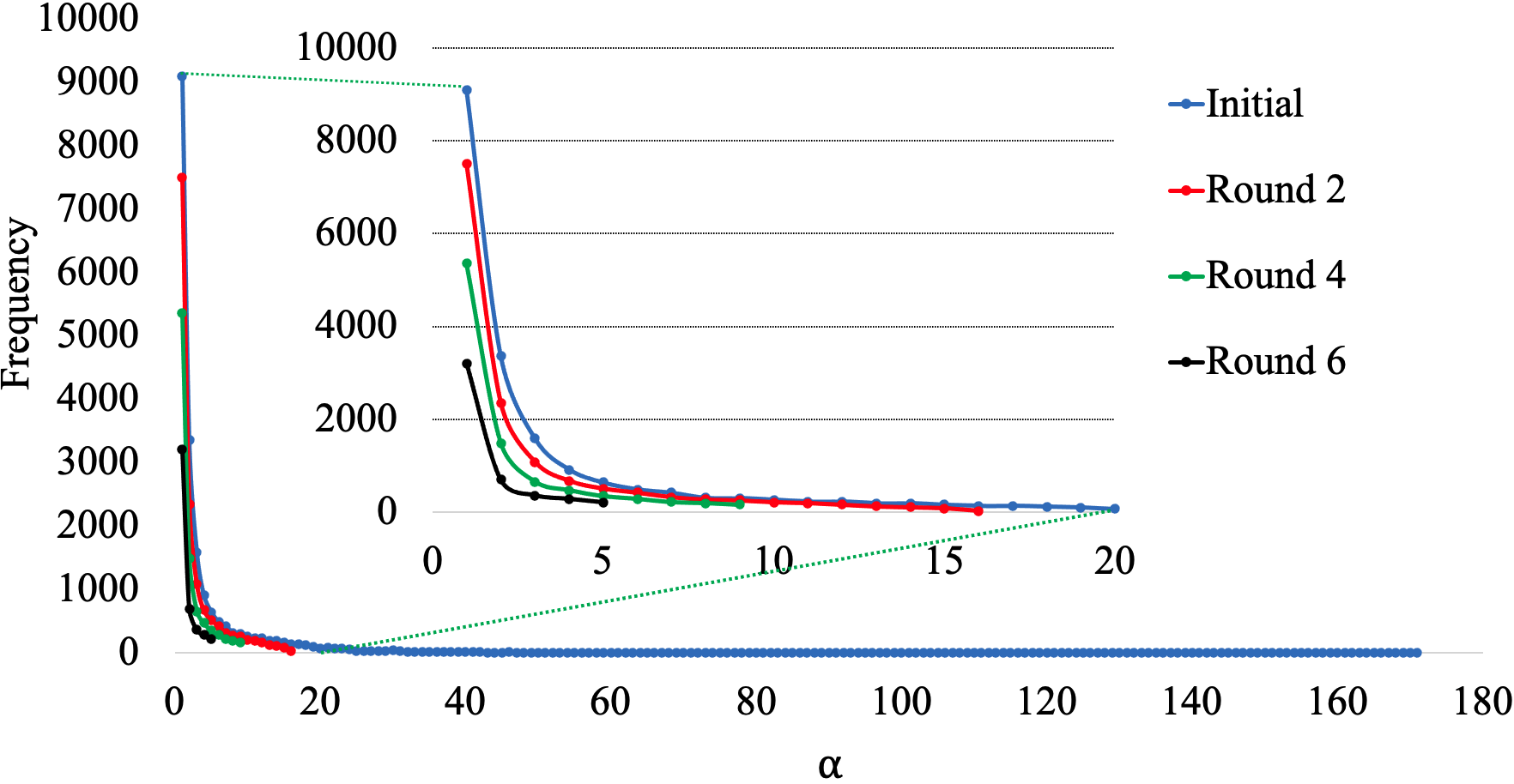}
\vspace{-3mm}
\caption{Histogram of $\alpha$ through various Rounds (Pubmed). The inset
shows a magnified view.}
\label{fig:alpha_histogram}
\vspace{-6mm}
\end{figure}

The effectiveness of the approach is illustrated in
Fig.~\ref{fig:alpha_histogram}, which shows the histogram of $\alpha_i$
distributions in the input buffer after each Round.  The initial distribution
corresponds to the power-law degree distribution, and in each successive Round,
the histogram grows flatter -- with both the peak frequency and maximum $\alpha$
becoming lower, thus mitigating the problems of power-law distribution.  In
contrast, HyGCN ignores the power-law problem, and AWB-GCN overcomes it using
high inter-PE communication.  Moreover, our approach is shown to be effective
even for much more intensive GAT computations (prior accelerators do not
address GATs).

\begin{figure}[t]
\centering
\hspace*{-0.1\linewidth}
\subfloat[]{\includegraphics[width=0.42\linewidth]{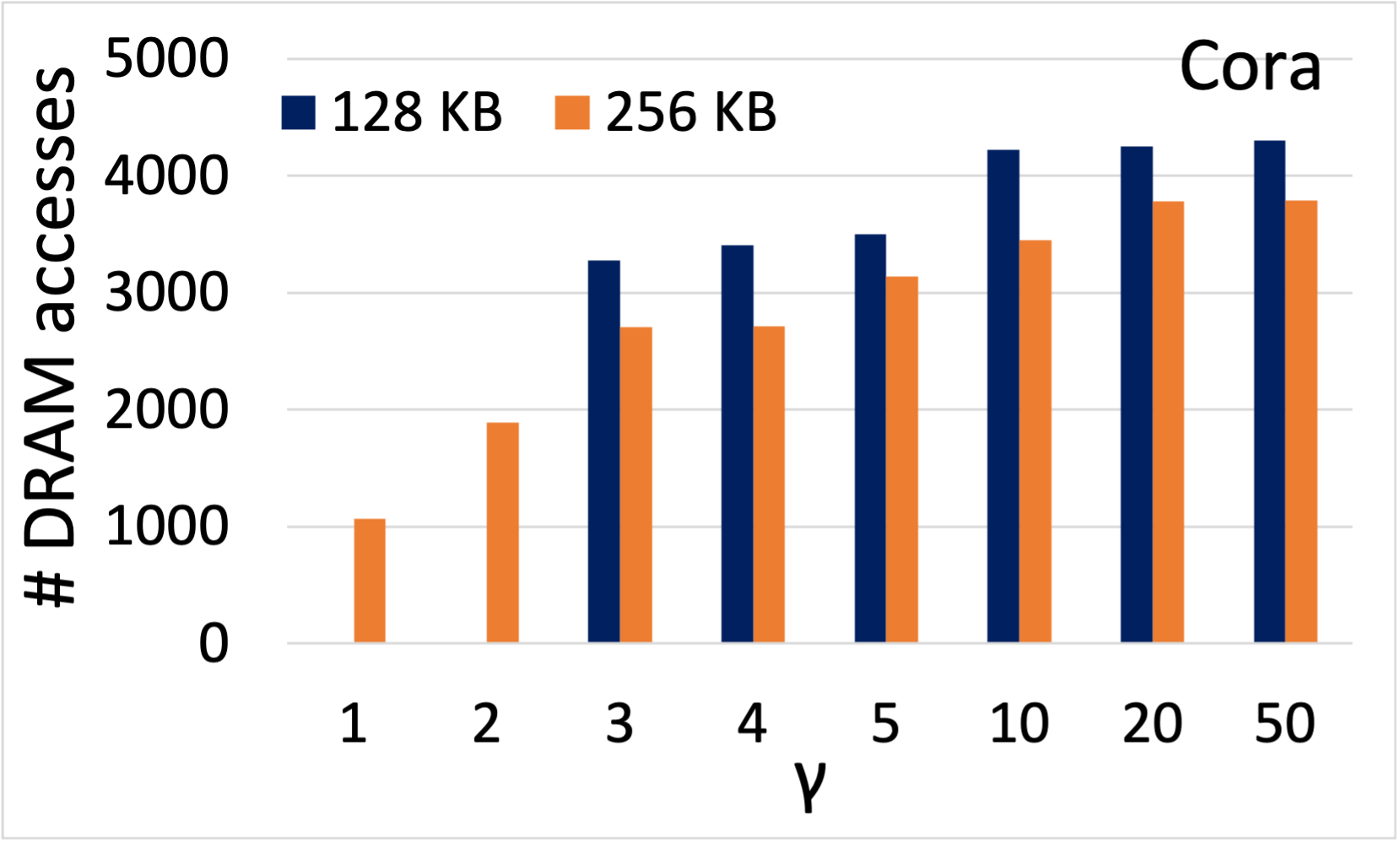}}
\subfloat[]{\includegraphics[width=0.39\linewidth]{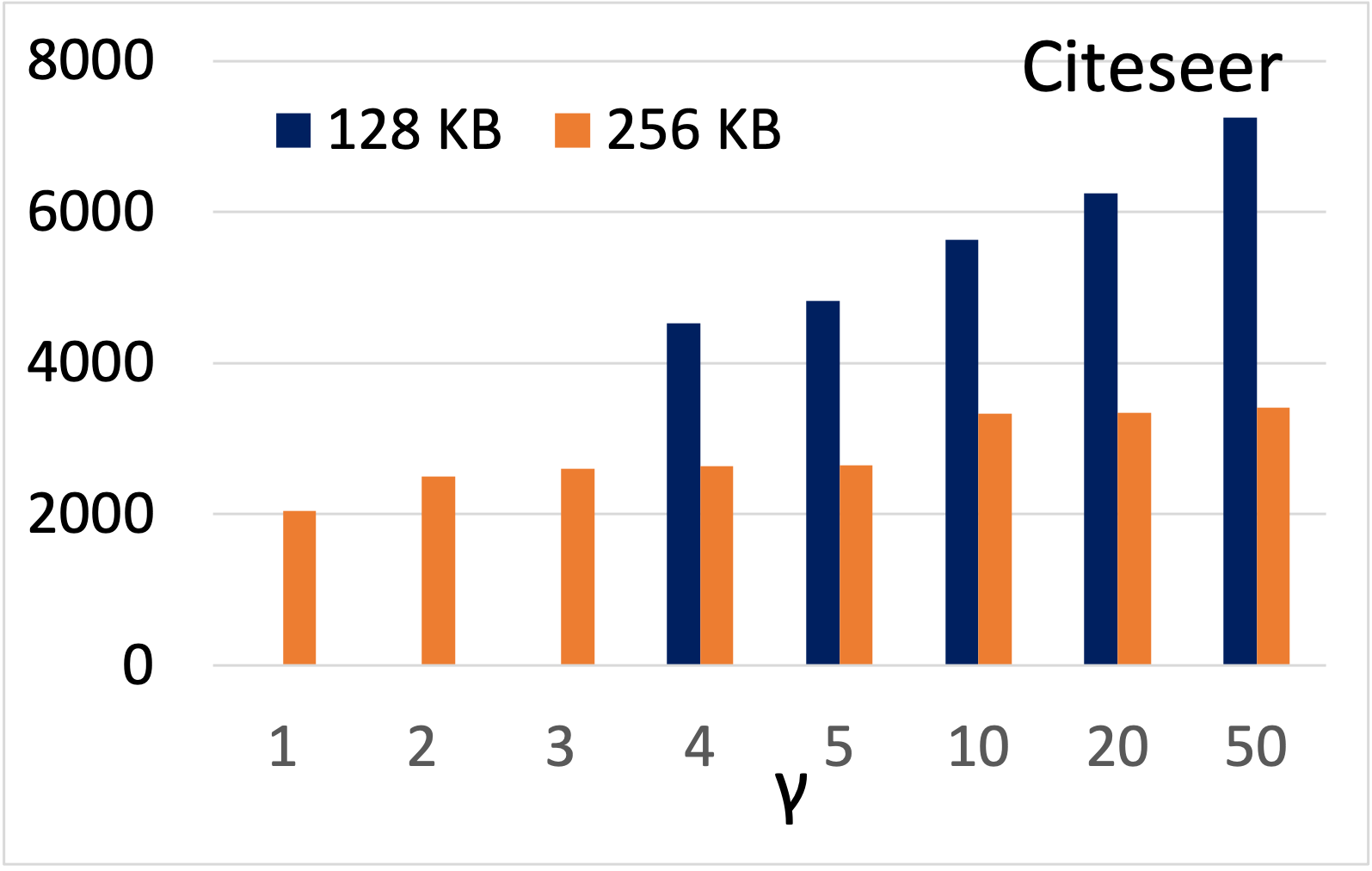}}
\subfloat[]{\includegraphics[width=0.39\linewidth]{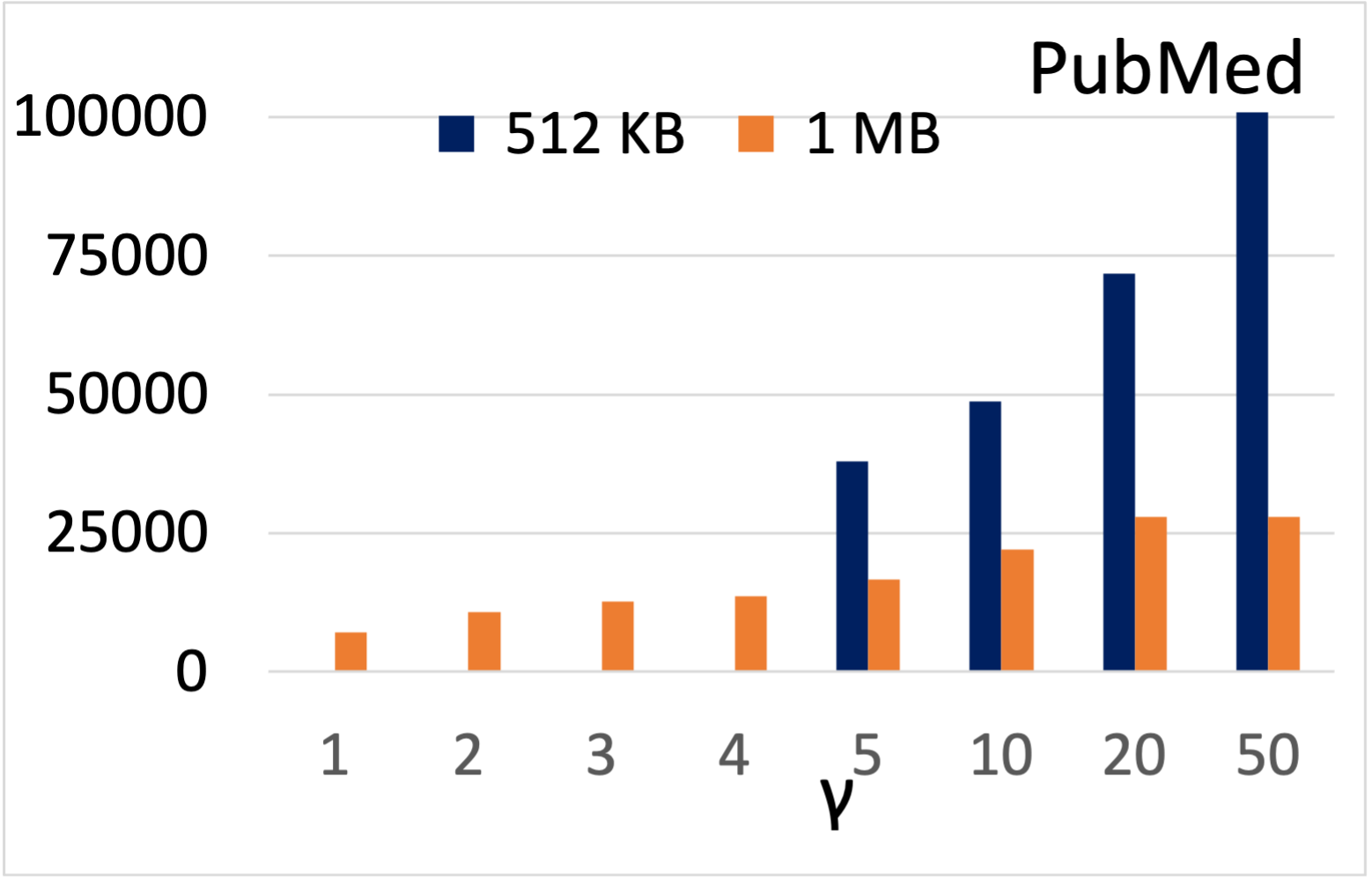}}
\vspace{-2mm}
\caption{Ablation study on $\gamma$: (a) Cora (b) Citeseer (c) Pubmed.}
\label{fig:Ablation}
\vspace{-6mm}
\end{figure}

Fig.~\ref{fig:Ablation} shows the impact of $\gamma$ on DRAM
accesses for three datasets.  As $\gamma$ increases, more vertices are evicted
and may have to be brought back to the cache, resulting in more DRAM accesses.
However, if $\gamma$ is too low, vertices may not be evicted from the cache,
resulting in deadlock as new vertices cannot be brought in. In our experiments,
we use a static value $\gamma=5$, but in practice, $\gamma$ may have to be
changed dynamically when deadlock arises. Deadlock detection is inexpensive and
is based on the number of total unprocessed edges in the partition, which is
monitored by a counter, and this dynamic scheme will be inexpensive in
hardware.

\vspace{-1mm}
\section{Related Work}
\label{sec:Related}

\noindent
There has been much work on CNN accelerators~\cite{EIE2016, chen2016eyeriss,
aimar2018nullhop, Jouppi2017, SCNN2017, BitFusion2018, Jouppi2020}, but these
are not efficient for processing GNNs.  Research on graph
analytics accelerations includes Graphicionado~\cite{ham2016graphicionado},
using modules tuned for irregular memory access, memory bandwidth bottlenecks,
and optimized on-chip memory utilization; FPGP~\cite{Dai16}, a multi-FPGA-based
accelerator that partitions the graph based on intervals and shards;
GraFBoost~\cite{jun2018grafboost}, employing a sort-reduce accelerator to
convert random accesses to flash memory to sequential ones;
GraphPIM~\cite{nai2017graphpim}, using Hybrid Memory Cube (HMC) for in-memory
processing.  However, graph accelerators target lightweight operations per
vertex, do not focus on data reuse, and would be challenged by computation-intensive
GNNs.

Software frameworks for GNNs include Deep Graph Library~\cite{wang2020deep},
AliGraph~\cite{yang2019aligraph}, and TensorFlow~\cite{tensorflow}.  Some
hardware accelerators have been proposed, but we know of no prior work that can
handle networks that require softmax nonlinearities on graphs such as GATs.
Although some GAT computations are addressed in~\cite{autenhardware}, the
crucial attention normalization step is left out.  To our knowledge, no methods
handle extreme input feature vector sparsity using graph-specific methods.

The HyGCN acclerator~\cite{yan2020hygcn} uses an Aggregation engine for
graph processing and a Combination engine for neural operations.  This requires
separate on-chip buffers for each engine: with workload imbalance at different
stages of computation, these are not fully utilized.
HyGCN must arbitrate off-chip memory access requests coming
from on-chip buffers of two different engines, which involves complicated
memory access control.  Using a single hardware platform optimized to handle
both the irregular graph computation and compute-intensive, albeit regular, DNN
computation, GNNIE achieves performance gains over HyGCN.  Moreover, HyGCN uses
sharding with window sliding/shrinking to reduce random memory access during
Aggregation: this has (1)~limited efficacy for highly sparse adjacency
matrices as the number of overlapping neighbors of vertices is a small farction
of the total number of vertices in a shard, undermining its efficacy; moreover,
no specific effort is made to address power-law degree distributions.
(2)~limited parallelism as the sliding window of the current shard depends on
the shrinking of the previous shard; HyGCN does not fully leverage data reuse
opportunities of high-degree vertices during Aggregation, performing
$(\tilde A \mathbf h_i^{l-1})W^l$, instead of cheaper $\tilde A(\mathbf
h_i^{l-1} W^l)$, which is much cheaper~\cite{Geng2020AWBGCNAG,liang2020engn}.
Moreover, input feature vector sparsity is not addressed and can result in
stalls, resulting in computational inefficiency. These factors
explain GNNIE's high speedups (35$\times$ on average) over HyGCN.

AWB-GCN~\cite{Geng2020AWBGCNAG}, which is limited only to GCNs and not general
GNNs, views the problem as a set of matrix operations.
It does not specifically try to reduce random memory accesses to the highly
sparse graph adjacency matrix.
Its dynamic scheduling scheme in AWB-GCN for workload redistribution among PEs
may incur high inter-PE communication, degrading energy efficiency.
EnGN~\cite{liang2020engn} uses a ring-edge-reduce (RER) dataflow for
Aggregation, where each PE broadcasts its data to other PEs in the same column. To
reduce communication, EnGN reorders the edges, but this is an energy-intensive
step, undermined by high sparsity in the adjacency matrix, that occurs
frequently as the limited number of cached edges are replaced.  The scheme has
large preprocessing costs. Although EnGN attempts dimension-aware stage
reordering to perform Aggregation or Weighting first, their results confirm
that in practice, the Weighting-first scheme used in GNNIE is better.

Prior accelerators have not fully explored load balancing.  Methods that
offload tasks to idler PEs (ring-edge-reduce~\cite{liang2020engn}, multistage
networks~\cite{Geng2020AWBGCNAG}) involve high communication and control
overheads.  GNNIE bypasses such approaches and uses the flexible MAC
architecture for load balancing, using heterogeneous PEs, and assigning
computation according to need. The idea is simple, effective, and easily
implemented. It results in low inter-PE communication, low control overhead,
and high speedup gain for the hardware overhead (Fig.  18). Preprocessing is
cheap and involves linear-time binning of vertex features blocks into groups.

Frequency-based caching techniques for graph data have been proposed
in~\cite{zhang2017making} using a programming interface.
However,~\cite{zhang2017making} is a purely software-based framework, agnostic
to the underlying hardware, for traditional graph analytics and uses a static
approach.  GNNIE uses a hardware-centric {\em dynamic} frequency-based caching scheme
that tracks the $\alpha$ value for each vertex with minimal hardware overhead,
and ensures serial access to DRAM.  Other schemes are also static and more
computational than GNNIE: they use hashing functions~\cite{Chen2021} or perform
more computation~\cite{Arai2016,Wang2021} in finding static
communities/partitions that do not specifically address cache size. On the
other hand, GNNIE's computationally cheap dynamic scheme automatically adapts
to the cache size using subgraphs built from vertices in the cache.
GRASP~\cite{faldu2020domain}, another cache management scheme for graph
analytics, employs a most-recently-used (MRU) approach. However, this scheme is
based on past history, while GNNIE's use of the unprocessed vertex count
measures future potential for a vertex.

\vspace{-1mm}
\section{Evaluation}
\label{sec:Results}

\noindent

\vspace{-2mm}
\subsection{Experimental Setup}
\label{sec:exptsetup}

\noindent
{\bf Accelerator Simulator:} We develop a cycle-accurate simulator to measure
the execution time in terms of the number of cycles required. The simulator
models each module of GNNIE and integrated with
Ramulator~\cite{kim2015ramulator} to model the memory access to the off-chip
HBM with 256 GB/s bandwidth.

Each module was implemented and synthesized in Verilog and the synthesized
design was verified through RTL simulations. Synopsys Design Compiler was used
to synthesize the accelerator at 32nm technology node with standard VT cell
library. The chip area, critical path delay, and dynamic/static power,
extracted from Design Compiler, are used for evaluating performance and energy.
CACTI 6.5 is used to estimate the area, energy consumption, and access latency
of on-chip buffers.  The energy of HBM 2.0 is 3.97 pJ/bit~\cite{o2017fine}.
The chip area is $15.6$mm$^2$ and its frequency is 1.3 GHz.

\begin{table}[!h]
\centering
\vspace{-3mm}
\caption{ Dataset Information~\cite{sen2008collective}}
\vspace{-3mm}
\label{tbl:dataset info}
\resizebox{\linewidth}{!}{
{\scriptsize
\begin{tabular}{|c|c|c|c|c|c|}
\hline
Dataset  &     Vertices   & Edges  & Feature Length  & Labels & Sparsity    \\ \hline

Cora (CR)       & 2708   &   10556  &   1433  &   7  &  98.73\%       \\ \hline
Citeseer (CS)         & 3327   & 9104  &   3703 &6    & 99.15\%           \\ \hline
Pubmed (PB)       & 19717   &   88648  &  500 & 3     &  90\%       \\ \hline
Protein-protein interaction (PPI)   & 56944  &   1.63M  &  50 & 121   &  98.1\%      \\ \hline
Reddit (RD)        & 232965   &   114.6M  & 602 & 41   &     48.4\%         \\ \hline
\end{tabular}
}
}
\vspace{-4mm}
\end{table}

\begin{table}[!h]
\centering
\caption{Convolution layer configurations (len[$\mathbf h_i^l$] = length of $\mathbf h_i^l$)}
\vspace{-3mm}
\label{tbl:Convolution layer configuration}
\resizebox{\linewidth}{!}{
{\scriptsize
\begin{tabular}{|l|c|c|c|}
\hline
GNN Model & Weighting                        & Aggregation & Sample size    \\ \hline

GAT       & len[$\mathbf h_i^l$] , 128       & Sum         & $--$          \\ \hline
GCN       & len[$\mathbf h_i^l$] , 128       & Sum         & $--$                \\ \hline
GraphSAGE & len[$\mathbf h_i^l$] , 128       & Max         & 25               \\ \hline
GINConv   & len[$\mathbf h_i^l$] , 128 / 128 & Sum         & $--$               \\ \hline
DiffPool ($\mbox {GCN}\textsubscript{pool}$)
          & len[$\mathbf h_i^l$] , 128       & Sum         & $--$               \\ \hline
DiffPool ($\mbox {GCN}\textsubscript{embedding}$)
          & len[$\mathbf h_i^l$] , 128       & Sum         & $--$               \\ \hline
\end{tabular}
}
}
\vspace{-3mm}
\end{table}

\noindent
{\bf Benchmark GNN Datasets and Models:}
For evaluation of the performance of GNNIE, we used the benchmark graph
datasets listed in Table~\ref{tbl:dataset info}. We used five GNN models for
evaluations, i.e., GAT, GCN, GraphSAGE, GINConv, and DiffPool. 
The convolution layer configurations are shown in
Table~\ref{tbl:Convolution layer configuration}.
All preprocessing costs are included in the evaluation. 

\noindent
{\bf Configurations for Baseline/Cross-Platform Comparison:}
We first compare GNNIE against two baseline architectures, i.e., a
general-purpose CPU and a GPU. The CPU platform is
equipped with Intel Xeon Gold 6132@2.60GHz and 768 GB DDR4. The GPU platform is
equipped with V100 Tesla V100S-PCI @1.25GHz and 32 GB HBM2. 

For GNNIE, the sizes of output and weight buffers are 1MB and 128KB,
respectively. The input buffer size is 256KB for the smaller datasets (CR, CS)
and 512KB for the larger datasets (PB, PPI, RD). The area and power numbers
reported later correspond to the larger input buffer size. 
The output buffer is larger since it must
cache many partial results before they are aggregated, particularly for
high-degree vertices.  For a 1-byte weight, for the dataset with the largest
feature vector ($\sim$4K for CS), to keep 16 CPE columns occupied, the buffer
size is 4K$\times$16$\times$2 (for double-buffering) = 128KB.

The dimension of the CPE array is $16 \times 16$ and consists of four MAC units
for CPE row number 1 to 8, five MAC units for CPE row number 8 to 12 and six
MAC units for CPE row number 13 to 16.  The number of MACs per CPE was chosen
through design space exploration, optimizing the cost-to-benefit ratio (speedup
gain~:~hardware overhead).


\vspace{-2mm}
\subsection{Baseline Platform Comparisons}

\begin{figure}[t]
\centering
\hspace*{-0.05\linewidth}
{\small (a)}\subfloat{\includegraphics[width=\linewidth]{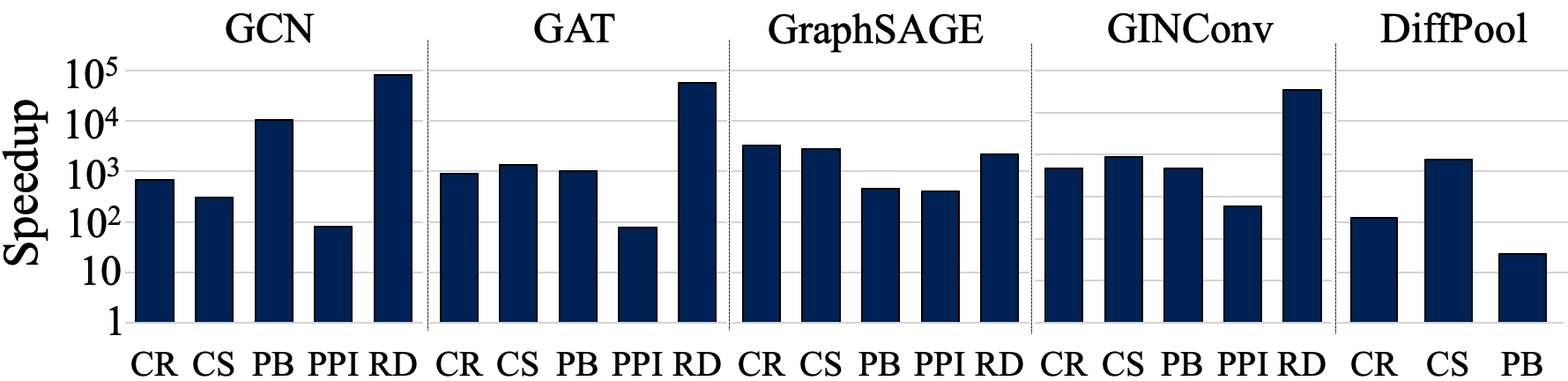}}

\vspace{-2mm}
\hspace*{-0.05\linewidth}
{\small (b)}\subfloat{\includegraphics[width=\linewidth]{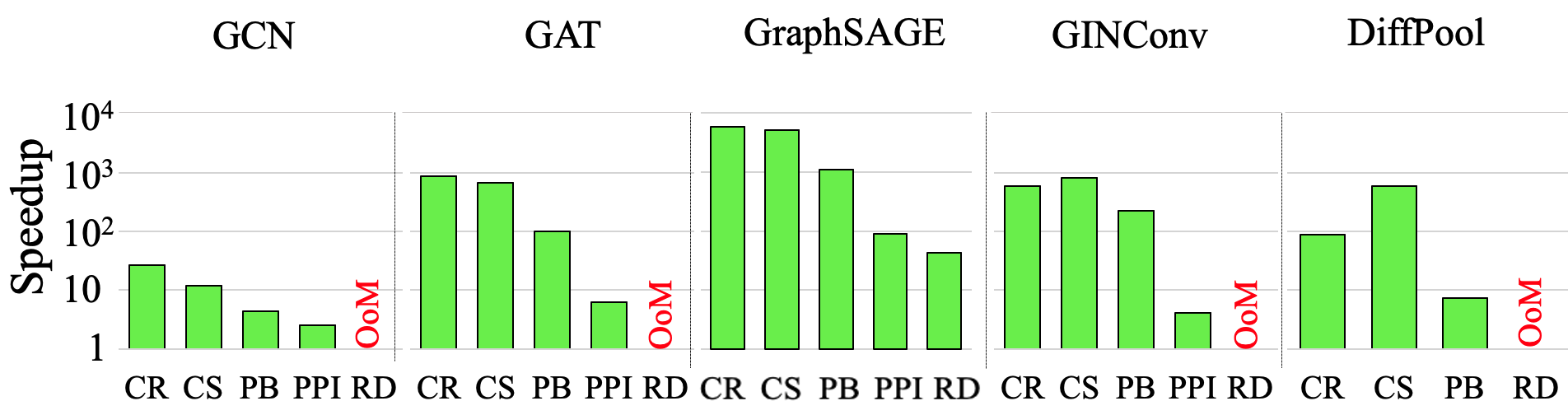}}
\vspace{-2mm}
\caption{GNNIE performance vs. (a) PyG-CPU (b) PyG-GPU.}
\label{fig:CPU GPU speed up}
\vspace{-6mm}
\end{figure}

\noindent
{\bf Performance comparisons with CPU and GPU:}
To make a fair performance comparison with the general-purpose CPU and GPU, we
implement the GNN models with the PyTorch Geometric (PyG) software framework.
The PyG-based implementations for CPU and GPU used in our experiment are
denoted as PyG-CPU and PyG-GPU, respectively. Neighborhood sampling for
GraphSAGE is based on cycling through a pregenerated set of random numbers: the
cost of random number generation is included in the evaluation.  As
shown in Fig.~\ref{fig:CPU GPU speed up}(a), the average speedup of GNNIE over
the PyG-CPU across the datasets used in our experiment for GCN, GAT, GraphSAGE,
GINConv, and DiffPool are 18556$\times$, 12120$\times$, 1827$\times$,
72954$\times$, and 615$\times$, respectively. According to
Fig.~\ref{fig:CPU GPU speed up}(b) the average speedup of GNNIE over the
PyG-GPU across the datasets used for GCN, GAT, GraphSAGE, GINConv, and DiffPool
are 11$\times$, 416$\times$, 2427$\times$, 412$\times$, and 231$\times$,
respectively.

The speedup calculations account for preprocessing
due to degree-based vertex reordering (Aggregation) and workload
reordering (Weighting).  For both, we use a binning approach that has linear
time complexity. For GraphSAGE, the time taken for sampling the vertex neighborhood 
is included.

The speedup comes from several GNNIE optimizations: (i)~The segmentation of
vertex feature vectors and their assignment in our FM architecture tackles the
feature vector sparsity challenge. (ii)~Our degree-aware cache replacement
policy avoids random memory accesses to DRAM; datasets with lower speedups
(e.g., PPI) have less strong power-law degree distributions and are unable to
fully benefit from our methods.  (iii)~During Weighting, distributing the
computation across multiple batches enables weight reuse, increasing
efficiency.  Note that PyG-CPU and PyG-GPU do not allow our dynamic caching
scheme to be implemented within their purely software based frameworks.

\vspace{-1mm}
\subsection{Cross-platform Comparisons}
\label{sec:xplatform}

\noindent
We conduct cross-platform performance comparisons with HyGCN and AWB-GCN.
Neither computes exponentiation for softmax, required by GATs, and
AWB-GCN only implements GCN. 
Thus, for GCN we perform a comparison with HyGCN and AWB-GCN.  For GraphSAGE
and GINConv we also show a comparison with HyGCN. Unlike the original
implementations, HyGCN uses 128 channels for hidden layers of all the GNN
models, and therefore we have also configured the hidden layers similarly
(Table~\ref{tbl:Convolution layer configuration}). 
To compare with HyGCN, AWB-GCN runs the customized GCN model with 128 channels
for hidden layers on a E5-2680v3 CPU with PyG and reports relative speedup and
inference latency. We leverage inference latency data from AWB-GCN for
our comparison.

\begin{figure}[!htb]
\vspace{-0.5em}
\centering
\includegraphics[width=\linewidth]{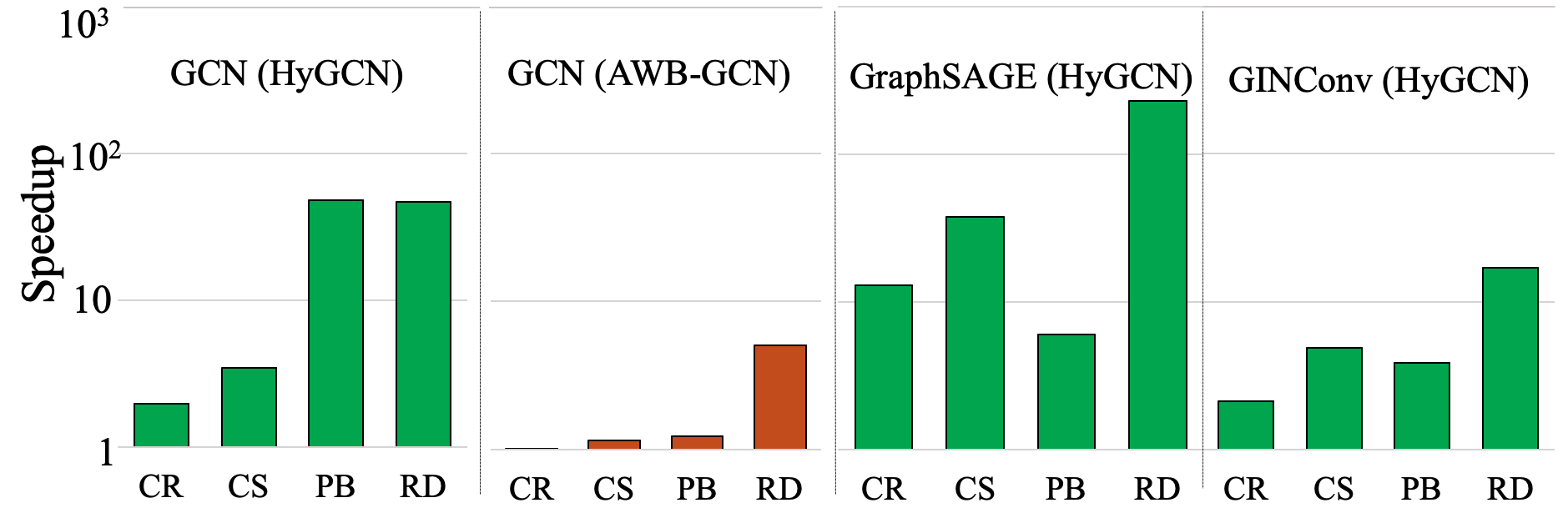}
\caption{Performance comparison with HyGCN and AWB-GCN.}
\label{fig:Cross plarform speed up}
\vspace{-2mm}
\end{figure}

To compute speedup over HyGCN for GraphSAGE, GINConv, and DiffPool we run the
GNN models on Intel Xeon Gold 6132@2.60GHz CPU, which has similar performance as
the E5-2680v3@2.50GHz CPU, and determine the relative speedup of GNNIE. We then
take a ratio of the computed relative speedup with the relative speedup of
HyGCN compared to E5-2680v3 CPU. Fig.~\ref{fig:Cross plarform speed up} shows
that for GCN, GNNIE achieves average speedup of 25$\times$ and 2.1$\times$
over HyGCN and AWB-GCN, respectively.  Compared to HyGCN, GNNIE achieves
average speedup of 72$\times$ and 7$\times$ for GraphSAGE and GINConv,
respectively. A comparison for DiffPool is not possible: HyGCN does not report
results on the widely used datasets that we evaluate.  As before, these speedup
comparisons include GNNIE preprocessing costs.

Even though the on-chip buffer size of HyGCN (24 MB + 128 KB) is much larger
than GNNIE (1.7 MB), GNNIE shows an average speedup of 35$\times$. The
improvements are attributable to differences between GNNIE and HyGCN
described in Sections~\ref{sec:Intro} and~\ref{sec:Related}.
Although AWB-GCN uses 4096 MACs against 1216 for GNNIE, GNNIE is 2.1$\times$ faster. 



\vspace{-2mm}
\subsection{Throughput and Energy Comparisons}

\noindent
Table~\ref{tbl:throughput} shows the throughput for various datasets for our
configuration of GNNIE. 
The table shows that the throughput degrades only moderately as
the graph size is increased.  

\begin{table}[b]
\centering
\vspace{-4mm}
\caption{Throughput for various datasets for GNNIE.}
\vspace{-3mm}
\label{tbl:throughput}
\resizebox{0.8\linewidth}{!}{
{\scriptsize
\begin{tabular}{|l|c|c|c|}
\hline
Peak      & Cora (CR) & Citeseer (CS) & Pubmed (PB) \\ \hline 
3.17 TOPS & 2.88 TOPS & 2.69 TOPS     & 2.57 TOPS   \\ \hline
\end{tabular}
\vspace{-5mm}
}
}
\end{table}

\begin{figure}[t]
\centering
\includegraphics[width=0.7\linewidth]{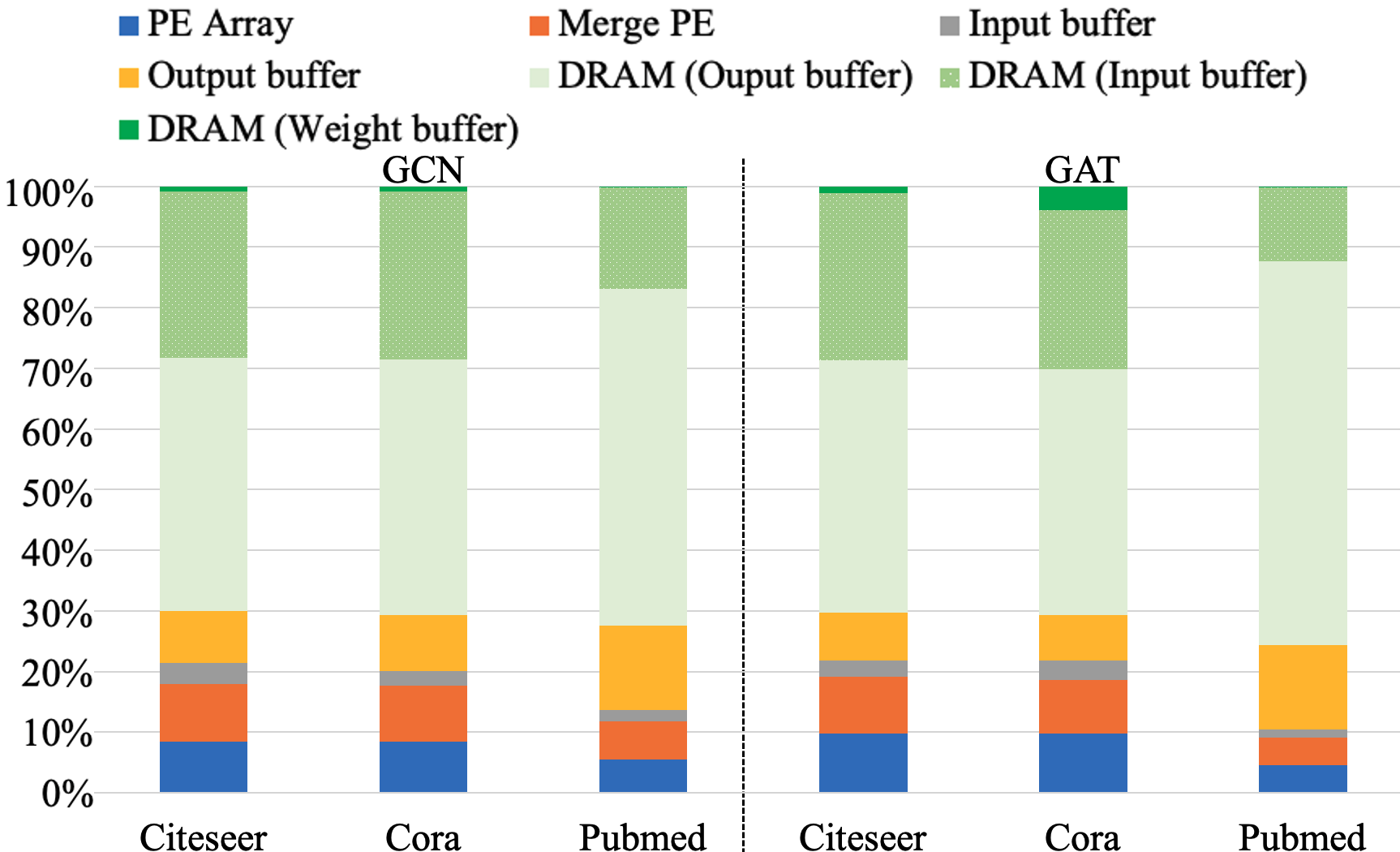}
\vspace{-2mm}
\caption{Energy breakdown for GCN and GAT.}
\label{fig:Breakdown of energy}
\vspace{-4mm}
\end{figure}

The power dissipation of GNNIE is 3.9W in 32nm, lower than HyGCN (6.7W
in 12nm), similar to recent CNN edge inference engines (Edge TPU,
Hailo-8, InferX1).  Fig.~\ref{fig:Breakdown of energy} shows the energy
breakdown for GNNIE for GAT and GCN across three datasets, including
DRAM energy required to supply the output, input, and weight buffers.  The
output buffer has the most of transactions with DRAM due to psum
storage. On-chip weight buffer energy is negligible and not shown.

\begin{figure}[t]
\centering
\includegraphics[width=0.57\linewidth]{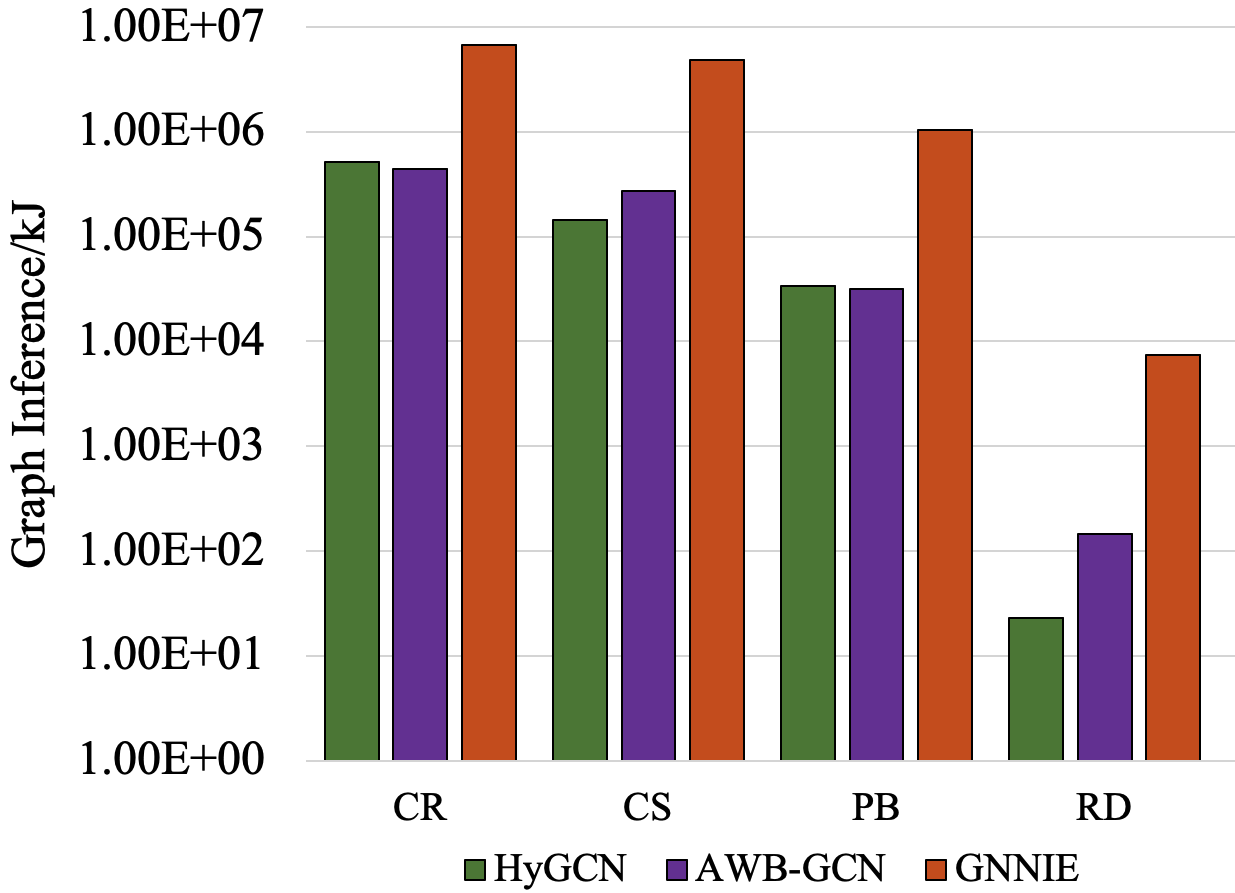}
\vspace{-2mm}
\caption{Energy efficiency: GNNIE vs. HyGCN, AWB-GCN.}
\label{fig:Energy efficiency comparison}
\vspace{-6mm}
\end{figure}

Fig.~\ref{fig:Energy efficiency comparison}
compares GNNIE's energy efficiency with prior works.
The efficiency ranges from
ranges from 2.3$\times$10$^1$ -- 5.2$\times$10$^5$ inferences/kJ for HyGCN and
1.5$\times$10$^2$ -- 4.4$\times$10$^5$ inferences/kJ for AWB-GCN.
GNNIE clearly outperforms the others,
going from 7.4$\times$10$^3$ -- 6.7$\times$10$^6$ inferences/kJ.

\vspace{-2mm}
\subsection{Optimization Analysis}

\noindent
We analyze key optimization techniques applied in GNNIE. To evaluate these
techniques we select a baseline design ({\bf Design A}) which uses four MACs
per CPE uniformly. Parameters for the flexible MAC architecture and on-chip
buffer sizes for all designs are as described at the end of
Section~\ref{sec:exptsetup}.  The dimension of the PE array in all cases is $16 \times 16$.

\begin{figure}
\centering
\vspace{-3mm}
{\small (a)}\subfloat{\includegraphics[width=0.97\linewidth]{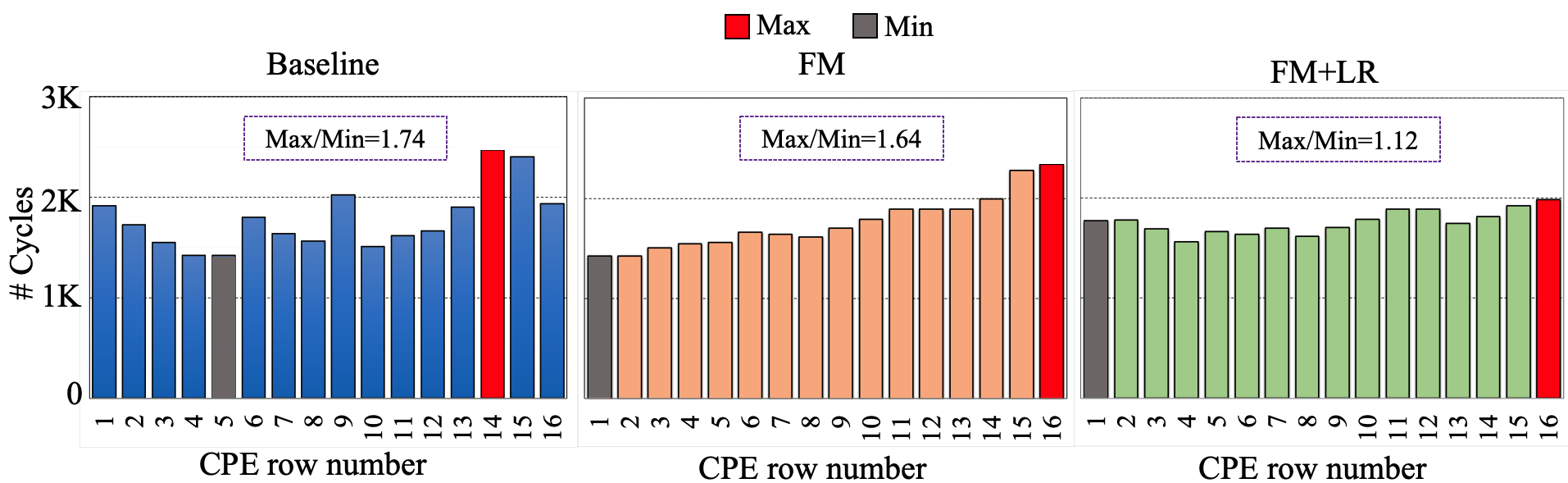}}\\

\vspace{-3mm}
{\small (b)}\subfloat{\includegraphics[width=0.97\linewidth]{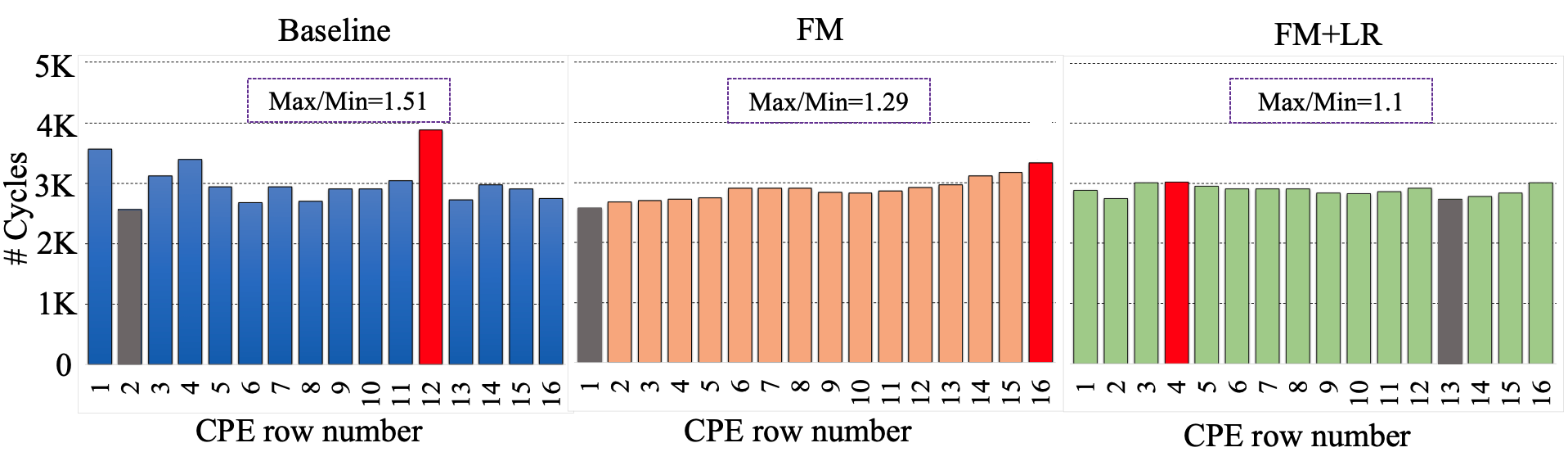}}\\

\vspace{-3mm}
{\small (c)}\subfloat{\includegraphics[width=0.97\linewidth]{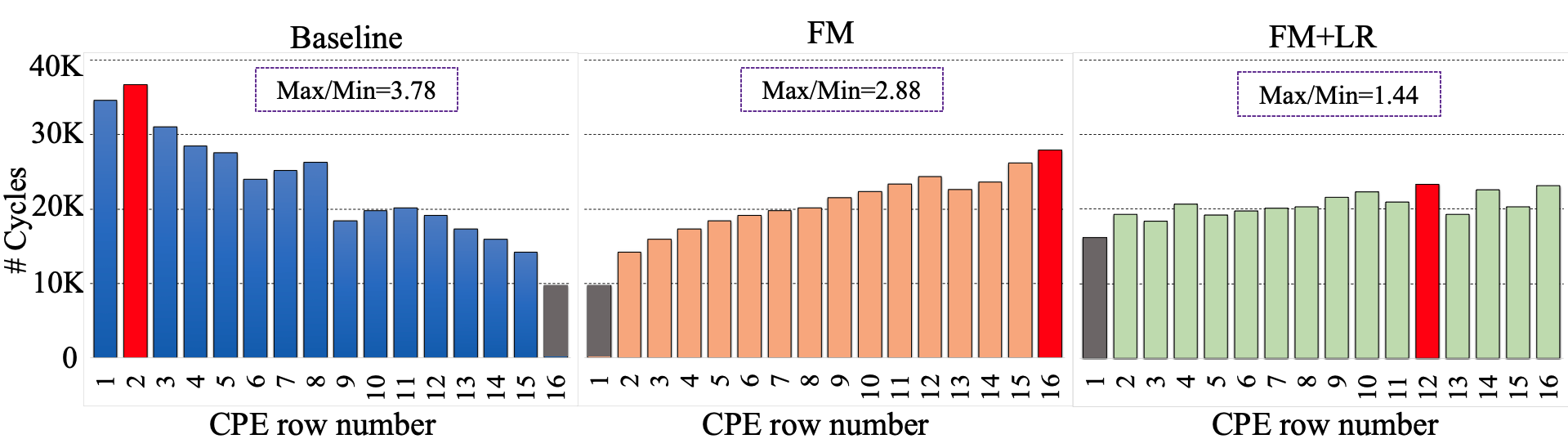}}
\vspace{-1mm}
\caption{CPE row workload in Weighting: (a) Cora (b) Citeseer (c) Pubmed.}
\label{fig:feature_transformation}
\vspace{-6mm}
\end{figure}

\noindent
{\bf Optimizing Weighting Time:}
We first analyze the performance improvement of applying flexible MACs
({\bf{FM}}) on the baseline design during Weighting.  For the Cora,
Citeseer, and Pubmed datasets, the workload distribution among the CPE rows for
the baseline (without load-balancing) and FM designs are shown in
Figs.~\ref{fig:feature_transformation}(a), (b), and (c), respectively. 
Due to vertex feature sparsity, the CPE rows in the baseline design
suffer from workload imbalance. The FM design smooths the
workload distribution among the CPE rows results in 6\% (Cora), 14\%
(Citeseer), and 31\% (Pubmed) reduction in the number of cycles required to
compute 16 elements of the output vertex features during Weighting. The imbalance
between the maximum and minimum is also reduced by FM.

For all datasets, the last four CPE rows require more cycles than others
(heavily loaded CPE rows) and the first four CPE rows finish computation
earlier (lightly loaded rows) in FM. We perform load redistribution ({\bf{LR}})
between ``LR pairs'' of heavily loaded and lightly loaded CPE rows,
offloading a portion of the workload from the heavily loaded CPE row to the
lightly loaded one.  The figure shows that applying LR on FM further smooths
the workload distribution, reducing the imbalance between the maximum and
minimum significantly, and also further reduces the number of cycles.

\ignore{
However, we need to judiciously select the workload being
offloaded so that it does not overload the lightly loaded CPE row. To do this
we analyze the nonzero workload assigned to heavily loaded CPE rows.
Fig.~\ref{fig:Workload histogram}(a) shows the histogram of nonzero workload
derived from vertex feature blocks assigned to the CPE row 16 for the Cora
dataset after applying FM. {\bf No ref to (b)?}

From the histogram we observe that, there are 604 vertex feature blocks which have only one 1 nonzero which is significantly lower than MACs in CPE row 16, i.e., 6. These vertex feature blocks do not get any significant speedup, however, they affect the total cycle count of CPE row 16. Thus we offload a portion of such workloads to one of the lightly loaded CPE rows which have a lower number of MACs than the heavily loaded CPE rows but enough MACs to finish computation of the offloaded workloads within a cycle.
}


\begin{figure}[b]
\vspace{-6mm}
\centering
\includegraphics[width=0.8\linewidth]{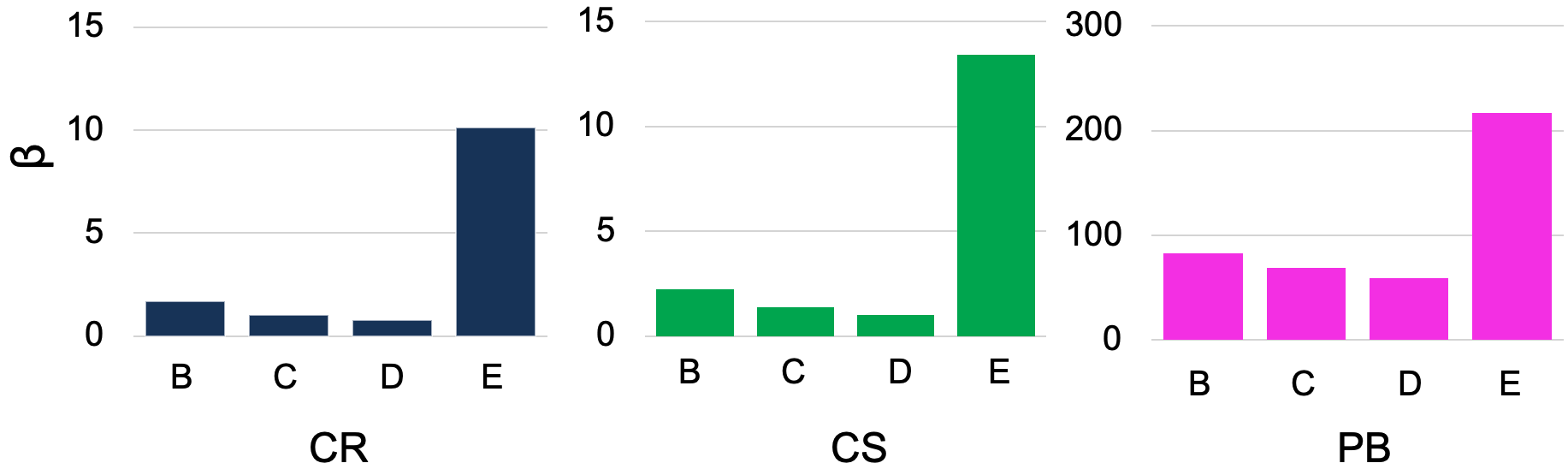}
\vspace{-1mm}
\caption{Speedup gain vs. hardware ratio for Designs B--E.}
\label{fig:MAC efficiency}
\end{figure}

\noindent
{\bf Speedup Gain vs. Hardware Overhead Ratio:} We introduce a metric, the
speedup gain vs. added hardware ratio:
\begin{equation}
\beta =\frac{\mbox{Baseline Design}  \mbox{ Cycles} - \mbox{Design[i]}  \mbox{ Cycles} }{\mbox{Design[i]}  \mbox{ MACs} - \mbox{Baseline Design}  \mbox{ MACs} }
\label{eq:MAC efficiency}
\end{equation}
where i refers to a specific design that is compared with the baseline.  We
measure $\beta$ with respect to the baseline with 1024 MACs (4 MACs/CPE). The
gain in speedup is measured in terms of reduction of the number of cycles
required for Weighting as MACs are increased in CPEs of the baseline design
uniformly. The additional hardware overhead is measured in terms of the number
MACs added to the baseline design. We compute $\beta$ for four designs with
respect to the reference designs. These design choices are as follows: (i)~5
MACs per CPE (\textbf{i.e., Design B}, 1280 MACs in all), (ii)~6 MACs
per CPE (\textbf{i.e., Design C}, 1536 MACs in all), (iii)~7 MACs
per CPE (\textbf{i.e., Design D}, 1792 MACs in all), (iv)~flexible
MAC architecture for GNNIE, described at the end of Section~\ref{sec:exptsetup}
(\textbf{i.e., Design E}, 1216 MACs in all).

Fig.~\ref{fig:MAC efficiency} plots $\beta$ on the three datasets used in our
experiment for the four design choices. As MAC units are added uniformly to the
baseline design $\beta$ drops and is lowest for Design D across all datasets.
$\beta$ drops for Designs B, C, and D as the high sparsity and sparsity
variation among vertex features yield low speedup gains as more MACs are added.
By employing MACs among CPE rows as needed, the FM approach tackles
input vertex feature sparsity, achieving high $\beta$ across all datasets.

\noindent
{\bf Optimizing Aggregation Time:}
Our baseline design has 4~MACs/row (no FM), no load balancing (i.e., no
degree-dependent load distribution in Aggregation), and no graph-specific
caching (i.e., vertices are processed in order of ID).

We first evaluate our degree-aware graph reordering and our proposed cache
replacement policy ({\bf{CP}}).  We measure the execution time of the baseline
during Aggregation with and without CP.  Fig.~\ref{fig:Results for optimization
techniques}(left) shows that CP reduces Aggregation time by 11\% (Cora), 35\%
(Citeseer), and 80\% (Pubmed). This is due to reduced random
off-chip memory accesses as more edges in a subgraph are processed under
degree-aware caching.
 
\begin{figure}[t]
\centering
\vspace{4mm}
\includegraphics[width=2.7 in]{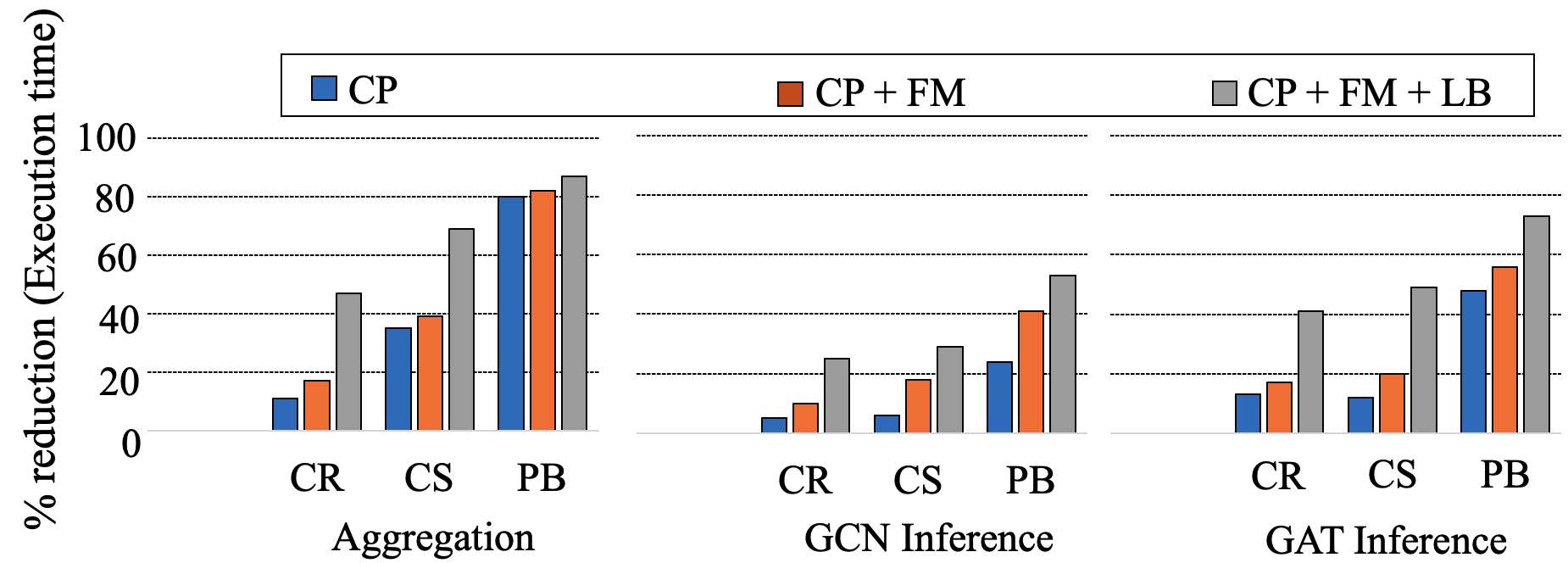}
\vspace{-2mm}
\caption{Effectiveness of GNNIE's optimization methods.}
\label{fig:Results for optimization techniques}
\vspace{-6mm}
\end{figure}

Next, we apply CP over FM to measure their combined effect.  From
Fig.~\ref{fig:Results for optimization techniques}(left), the added MACs in CP
+ FM yield gains of 17\% (Cora), 39\% (Citeseer), and 82\%
(Pubmed).

We add our approach for load-balancing ({\bf LB}) during Aggregation,
using the load distribution approach in Section~\ref{sec:map_edge}, on top of
CP+FM. The combined effect (CP+FM+LB) is shown in Fig.~\ref{fig:Results for
optimization techniques}(left) to reduce Aggregation time cumulatively by 47\%
(Cora), 69\% (Citeseer), and 87\% (Pubmed).

\noindent
{\bf Optimizing Inference Time:}
We evaluate our techniques on GCN and GAT inference time.  We first analyze the
effect of CP on inference time. Next, we incrementally add FM and LR
optimization to CP and measure their combined effect on inference time.
Finally, we add all load-balancing (LB) methods: the LR technique for Weighting
as well as load distribution during Aggregation.  Figs.~\ref{fig:Results for
optimization techniques}(middle) and (right) shows the reduction in the GCN and
GAT inference time, respectively for CP, CP+FM, and CP+FM+LB. The  reduction in
inference time is higher for Pubmed (19717 vertices) than Cora (2708 vertices),
indicating the scalability of GNNIE.

\vspace{-1mm}
\section{Conclusion}
\label{sec:Conclu}

\noindent
We have proposed GNNIE, a versatile GNN acceleration platform for to
a wide degree of GNNs, including GATs.  GNNIE efficiently works with unstructured
data, input vertex feature vector sparsity, and adjacency matrix sparsity, and
``power-law'' vertex degree distribution.  It mitigates load balancing issues,
computational bottlenecks, and irregular/random data accesses 
using multiple methods: splitting the computation into blocks to
leverage sparsity; optimized caching strategies; employing a flexible MAC
architecture in the CPE array. Substantial improvements over prior work are shown.

\clearpage
\bibliographystyle{IEEEtranS}
\bibliography{bib/main}

\end{document}